# FPA-FL: Incorporating Static Fault-proneness Analysis into Statistical Fault Localization

**Farid Feyzi, Saeed Parsa**[*]

Department of Computer Engineering, Iran University of Science and Technology, Tehran, Iran

Farid_feyzi@comp.iust.ac.ir, Parsa@iust.ac.ir

**Abstract**

Despite the proven applicability of the statistical methods in automatic fault localization, these approaches are biased by data collected from different executions of the program. This biasness could result in unstable statistical models which may vary dependent on test data provided for trial executions of the program. To resolve the difficulty, in this article a new 'fault-proneness'-aware statistical approach based on Elastic-Net regression, namely FPA-FL is proposed. The main idea behind FPA-FL is to consider the static structure and the fault-proneness of the program statements in addition to their dynamic correlations with the program termination state. The grouping effect of FPA-FL is helpful for finding multiple faults and supporting scalability. To provide the context of failure, cause-effect chains of program faults are discovered. FPA-FL is evaluated from different viewpoints on well-known test suites. The results reveal high fault localization performance of our approach, compared with similar techniques in the literature.

## 1. Introduction

In this article, a hybrid automated fault localization method is introduced that takes advantage of the merits of both the static and dynamic analysis. In fact, statistical methods are biased by the test data used to run the program under test (PUT). The biasness could be alleviated by considering the static structure of the program while modeling dynamic behavior of the program to highlight error prone regions of the program. Existing statistical approaches [1-9] are merely based on program execution data and to the best of our knowledge, no method has gained benefit from knowledge on fault-proneness associated with different portions of the code. Studies have shown that certain structures of code such as the depth of the nested loops and the modified or recently developed parts of the code are more likely to be faulty [10-11]. Generally speaking, code complexity correlates with the defect rate and robustness of the application [11-13]. Code which is too complex is often the reason for bad code quality and erroneous programs. Static code metrics are direct measurements of source code and may potentially relate to code quality, and therefore to fault-proneness [12]. The fault-proneness analysis is especially beneficial for large programs, where searching for likely failure origins could be accomplished more effectively if programmers give more attention to fault-prone portions and scrutinize them thoroughly. We have applied a defect prediction method, based on code complexity metrics, to estimate the static fault-proneness likelihood (FPL) of methods for subject programs. To reduce the high false positive rate of static methods [14-15], the estimated value is further refined, considering the number of appearance of the involved statements in passing executions. It should be noted that coincidental correctness occurs frequently [16-18] and could negatively affect the refinement process. To differentiate these tests a clustering-based approach [18] is employed. The refined FPLs are applied to improve the suspiciousness scores obtained from dynamic analysis methods. We have used the refined FPLs to adjust the correlation of the statements to be selected as bug predictors in an Elastic-Net regression.

In a typical program, the statements could be highly correlated with each other. The correlation caused by control and data dependencies amongst the statements represented as the predictor variables does not let us apply the ordinary methods to solve the regression equation, precisely. Typically, penalized regression methods have the capability to cope with correlated predictor variables [19]. Therefore, a penalized logistic regression method with a feature selection capability could be advantageous to identify fault relevant statements, while considering the correlation amongst the statements. The difficulty is to preserve the feature selection capability by penalizing the coefficients of the predictor variables while considering the structural dependencies amongst the statements represented as the predictor variables in the regression model. The proposed statistical approach builds a regularized logistic regression model based on fault candidate program statements.

The regression predictors are initially selected as those statements appearing in dynamic slices of the statements revealing the faulty results. Using a combination of the statistical approach to fault localization and backward slicing it is possible to take advantage of the capability of statistical approaches to rank statements according to their fault suspiciousness and the strength of backward dynamic slicing in restricting the statements to those included in the cause-effect chain of the failure(s). The impact of each of the statements could be appropriately estimated by computing the coefficient of its corresponding predictor variable in the regression model. Besides, the simultaneous impact (i.e., combinatorial effect) of correlated program statements on the program termination state could be estimated by using a regression-based statistical approach. Considering this fact is of great importance because the most program failures are only revealed when a specific combination of correlated statements are executed [20].

A significant challenge for statistical debugging techniques is to find multiple bugs in programs. It has been shown that single-bug algorithms are not effective for multiple-bug cases [1][21-22]. The grouping effect of the fault suspicious statements, appearing as predictors in the Elastic-Net regression, could reveal multiple bugs in a program. In fact, each group of statements, appear in execution paths with a relatively high number of common sub-paths, often include a distinct faulty statement in a multiple fault program. This property is helpful for finding multiple faults and correlated predictor variables.

Another issue known as $P>>N$ problem [23] may occur when the debugging method relies on modeling all program statements in a linear model such as regression [24]. This problem happens when the number of statements is much bigger than those of the executions. Employing a penalized regression method with grouping effect [23] enables us to handle this problem in large-sized programs.

The main contributions of this article, which include extensions of previous works [25-26][73], are as follows:
1) Considering the complex interactions among program statements and analyzing their effect on each other and on program termination status by using an Elastic-Net regression based method.
2) Handling the known $P>>N$ problem by using a penalized regression method that makes the proposed approach scalable.
3) Considering the program structure in statistical fault localization. FPA-FL attempts to discover the location of faults by taking into account the program structure and static fault-proneness analysis.
4) Finding multiple faults in programs. Grouping effect of Elastic-Net is very useful in dealing with this problem.

The remaining part of the paper is organized as follows. In Section 2, we first provide an overview of FPA-FL and then describe its steps in detail. This section describes how the execution data is used to construct expanded backward dynamic slices (EBDS) and then is converted to program spectra. Clustering execution vectors and identifying likely coincidentally correct tests are also described in this section. The experiments and results are shown in Section 3. Some discussions containing the related works and threats to validity are presented in Sections 4 and 5. Finally, the concluding remarks are mentioned in Section 6.

## 2. FPA-FL method overview

In this section, the main idea behind our proposed fault localization technique is presented. As shown in Figure 1, FPA-FL has four main stages. These stages include computing backward dynamic slice (BDS), identifying coincidentally correct tests, estimating static fault-proneness likelihood and finally localizing faults. The initial inputs are a buggy program and a test suite. These stages are detailed in the following sub-sections.

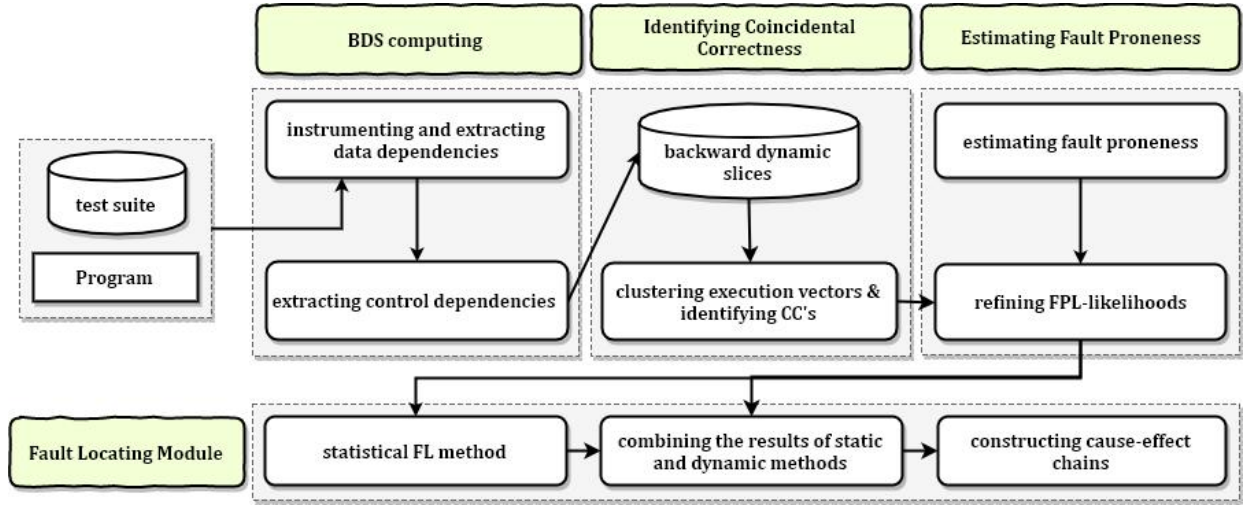

Figure 1. The framework of FPA-FL

### 2.1 Computing BDSs and constructing execution vectors

In order to minimize the search space in Spectrum-based Fault Localization (SBFL) approaches, we suggest the use of expanded BDSs, instead of all the program statements, to build the program spectra. Since a failure in a failing run is mostly manifested as a wrong output value, the BDS of the value frequently captures the faulty code is responsible for producing the incorrect value [14]. Therefore, the first step in computing BDSs and constructing execution vectors is to analyze the program and recognize the output statement(s) to find out which ones generate undesired results. Some programs may have multiple outputs and the debugger should identify the one producing incorrect values. The main concern is how to consider data and control dependence among the output value and the faulty code when analyzing different failing and passing tests. Since we do not know the location of faulty code, we compute the backward full dynamic slice starting from the incorrect value which frequently captures the faulty statement(s) and later we expand the BDS to include more suspicious statements according to our previous work in [2]. To compute EBDSs, we calculate two conditional probabilities to find out whether given statements, which are executed but are not included in BDSs, have a potential to affect the program result. After computing EBDSs, we convert each execution to a vector where each element of the vector corresponds to a program statement. Then, we give values to the elements of the vector according to the presence or absence of the element in the EBDSs of corresponding execution.

To compute BDSs we used the dynamic slicing framework introduced in [27]. The framework instruments a given program and executes a GCC compiler to generate binaries and collects program dynamic data to produce its dynamic dependence graph. The framework contains two main tools: Valgrind and Diablo. The instrumentation is done by Valgrind memory debugger and profiler who also identifies the data dependence among statement execution instances. The Diablo tool is capable of producing control flow graph (i.e. the control dependence among statements) from the generated binaries. We also used WET tool to compute BDSs from an incorrect output value.

## 2.2 Clustering execution vectors and identifying likely coincidentally correct tests

In fact, EBDSs of program executions are computed with the aim of determining potentially faulty statements which will be used as covariates in our Elastic-Net regression model. Based on this idea and assuming that existing passing and failing executions for program $P$ are presented as two failing and passing labeled execution vectors in program space, $Prog\text{-}Space(P)$, we have applied $k$-means clustering method. The execution points in each cluster share common executed and slice included statements and therefore they are approximately related to a particular program execution path.

Given a set of $m$ execution vectors for program $P$, without regarding the fail\pass label of the vectors, $k$-means aims to partition them into $k$ sets ($k \leq m$) $CL = \{cl_1, cl_2, \ldots, cl_k\}$ according to the following within-cluster function which minimizes the sum of squared Euclidean distances:

$$\underset{CL}{\mathbf{argmin}} \sum_{i=1}^{k} \sum_{V_k^P \in cl_i} \|V_k^P - c_i\|^2 \qquad (1)$$

Where $c_i$ is the center of execution vectors in cluster $cl_i$ and $V_k^P$ is an execution vector corresponding to $tc_k$ in $Prog - Space(P)$. The simple $K$-means takes the number of clusters as a parameter. In our experiment, similar to [21], this number is set according to the size of test suite, $T$. Let $CN$ denote the number of clusters, $CN = |T| * p$, where $|T|$ is the size of $T$ and $0 < p < 1$. The previous studies worked well in the cases of $p = 2\%$ to $p = 5\%$. Hence, we set $p = 1\%, 2\%, 4\%$ and $6\%$ in our experiment, respectively.

In general, there is no a fixed ratio of clustering (i.e. the number of clusters) suitable for all cases [18]. It mainly depends on the size of the test suite, and how much risk the developers are willing to take in order to identify the coincidentally correct tests. If the size of test suite is large, a relatively low value of $p$ can be chosen to keep the level of the false negatives low. If the developers have a high demand for accuracy of the recognition of coincidental correctness, a relatively high value of $p$ can be chosen to keep the level of the false positives low. After clustering the execution vectors, we use the technique proposed in [2] to expand BDS to the suspicious statements of the relevant slice [28] without computing the relevant slices, directly, to reduce the computational cost and decrease the number of irrelevant statements.

We leverage a clustering-based strategy to identify the subset of a test suite that is possible to be coincidentally correct [18]. coincidental correctness occurs when a test case executes the faulty elements but no failure is triggered. Previous studies have demonstrated that coincidental correctness is prevalent [16-17]. When coincidentally correct tests are present, the faulty elements will likely be ranked as less suspicious than when they are not present. As shown in the previous studies [17][29], the efficiency and accuracy of SBFL can be improved by cleaning the coincidentally correct tests. However, it is difficult to identify coincidental correctness because we do not know the locations of faulty elements in advance.

After clustering of the BDSs of executions, passing tests which are grouped into the same cluster with the failing ones are very likely to be coincidentally correct. Coincidentally correct test traces are usually expected to be more similar to failing executions because they analogously executed faulty statement(s). So, clusters that only include passing executions can be considered with high confidence as 'coincidental correctness'-free and can be employed in the refinement process of static FPLs.

## 2.3 Fault-proneness analysis based on static code metrics

Programming experiences have shown that faults distribution across programs is not uniform and some parts of programs are more likely to contain faults. In general, fault-proneness of a part of a program is directly related to its complexity. For instance, the nested function calls are more likely to be faulty compared with simple function calls. A programmer who writes a recursive function is more likely to make mistakes compared with a programmer who writes non-recursive function. Those parts of a program which are developed or modified more recently are more likely to be faulty [30-31]. So, there exists a relationship between the measures of software complexity and the faults found during testing and operation phases. These are properties that may potentially relate to code quality, and therefore to fault-proneness.

It has been shown that the distribution of faults over a system can be modeled by a Weibull probability distribution [32]. This motivates the use of software fault prediction models which provide an upfront indication of whether a portion of code is likely to contain faults, i.e., is fault-prone. Such prediction models are especially beneficial for large-scale systems, where the search space for bug localization is comparatively large. However, there has been considerable debate about the extent to which software fault prediction models constructed from code complexity metrics [33-34] actually contribute to supporting software testing processes. More recently, the validity of software fault prediction using static code features has been empirically illustrated by, e.g., Menzies et al., who stated that static code features are useful, easy to use, and widely used [35]. This observation was later also confirmed by other work, e.g., [36]. Static code features are known to be correlated. Authors in [37], investigated different static code features and claimed that the metrics could be grouped into four categories. A first category related to metrics derived from flowgraphs (i.e., McCabe metrics), while a second category contained metrics related to the size and item count of a program. The two other categories represented different types of Halstead metrics. This motivates the use of a feature selection procedure, especially when applying techniques that do not include some sort of embedded feature selection. When analyzing static code metrics, it is important to know the level of granularity at which they were captured. Common granularities include the file and package level, as well as the module level. In this article, the term module is used as a generic term to refer to the method level in C and Java programs.

Researchers have adopted different techniques to construct software fault prediction models. These include various statistical techniques such as logistic regression and Naive Bayes which explicitly construct an underlying probability model [38]. Furthermore, different machine learning techniques such as decision trees, support vector machines, and techniques that do not explicitly construct a prediction model but instead look at a set of most similar known cases have also been investigated [38]. We construct different prediction models according to [39] and apply them to predict fault-prone areas of the subject programs. Consequently, a static FPL for each program statement is obtained. Regarding the fact that many fault-prone elements of a program may not be related to the failure, we suggest refining FPLs based on coverage information.

### 2.3.1 Refining Static Fault-proneness Likelihoods Based on Coverage Information

Considering the fact that static FPLs are computed based on static program structure and code complexity metrics, the high false positive rate is inevitable. By false positive, we mean that there may be statements with high static FPLs that have no relation to the program failure. Not taking into account this fact may lead bias in the final ranking of statements. In other words, failure origins are not necessarily located in fault-prone portions of a program. So, before incorporating the estimated static FPLs of program statements in our statistical fault localization method, we propose to leverage coverage information to refine them.

With this understanding, we now explain how to refine the static FPL of a statement (say $x$) based on its appearance frequency in passing executions. If we consider that after identifying and separating likely coincidentally correct tests, there remain $k$ passing test cases and statement $x$ appears in $c$ out of them, the proposed refinement process is shown in equation (2).

$$\textbf{Refined static } FPL(x) = \begin{cases} \frac{|k| - |c|^\theta}{|k|} \times static\ FPL(x) & if\ |c|^\theta < |k| \\ 0.1 & if\ |c|^\theta \geq |k| \end{cases} \quad 1 < \theta \leq 2 \quad (2)$$

where $\theta$ is a constant between 1 and 2. More discussion on the impact of $\theta$ on the effectiveness of our proposed fault localization can be found in Section 3.3. Since we first attempt to separate likely coincidentally correct tests in a test suite and exclude them from the refinement process, it is expected to remain a small number of such tests in the test suite. Therefore, the refinement process could be done with confidence.

It is argued that often about 20% of the code is responsible for the 80% of program faults [40]. So, in our experiments, about 20-30 percent of program statements are considered as fault-prone and their static FPLs refined based on proposed refinement strategy. Static FPL of other statements is considered as zero and their suspiciousness scores are estimated using the Elastic-Net regression method.

## 2.4 Statistical fault localization Based on Elastic-Net Regression Model

The major difficulty with existing SBFL techniques is that they analyze each program statement in different failing and passing spectrum, in isolation. In other words, the suspiciousness score of a program statement is estimated without considering the combinatorial effect of other statements on program termination state. To address this problem, a new logistic regression method dedicated to automatic fault localization is introduced.

In this regard, for each failing test, the EBDS of the incorrect output value(s) is computed. Each failing slice contains statements which have data or control dependence with program's incorrect output value(s). We make a fault candidate set, $CS_F$, which contains statements included in the union of EBDS' of all failing tests. The candidate set is used to build a $m * n$ spectra matrix, the so-called *slice coverage matrix* (SCM), where $m$ and $n$ are the number of test cases and statements in the candidate set, respectively. The $n$ predictor vectors in addition to the first column of one's[1] make the columns of SCM. Each row $i$ of the matrix corresponds to program spectrum of the $i^{th}$ test case and each column $j$ corresponds to a particular statement in the candidate set of program. Each element $d_{ij}$ of the SCM, except those in the first column, depicts the number of times statement $j$ is executed in the $i^{th}$ execution. We have also a $m * 1$ pass/fail vector where each $y_i$ element of the vector is either 0 or 1, indicating failing or passing termination state of the program for the $i^{th}$ test case in the test suite. Note that although we build the columns of the SCM according to the statements in EBDS of the failing tests, the rows of the matrix contain both failing and passing test cases of the program.

**Definition 1.** $o_f = \{s_j |$ the output of statement($s_j$) is unexpected under some test cases$\}$ is a set of statements which outputs mismatch the expected output.

**Definition 2.** $EBDS(o_i, t_{f_j})$ is a function that returns the expanded backward dynamic slice on statement $o_i$ under test case $t_{f_j}$, where $o_i \epsilon o_f$ and $t_{f_j} \epsilon T_f$

Let $CS_F$ denotes the union of $EBDS(o_i, t_{f_j})$, $o_i \epsilon o_f$, $t_{f_j} \epsilon T_f$. In fact, $CS_F$ can be denoted by a set of statements $\{s'_1, s'_2, ..., s'_n\}$ shown in SCM of equation (3).

$$\underbrace{\begin{pmatrix} & s'_1 & s'_2 & \cdots & s'_n \\ \mathbf{1} & d_{1,1} & d_{1,2} & \cdots & d_{1,n} \\ \cdots & & \vdots & \ddots & \vdots \\ \mathbf{1} & d_{m,1} & d_{m,2} & \cdots & d_{m,n} \end{pmatrix}}_{Slice\ Coverage\ Matrix} \times \begin{pmatrix} \beta_0 \\ \vdots \\ \beta_n \end{pmatrix} = \begin{pmatrix} y_1 \\ \vdots \\ y_m \end{pmatrix} \quad (3)$$

where the elements $d_{ij}$ denote the number of times the statement $s'_j$ is executed in the $i^{th}$ test case.

Now, the aim is to estimate the coefficient of each statement to identify its impact on the program termination state. Logistic regression is an appropriate method to model the interrelations among statements and failing or passing status of a program. Regression analysis answers questions about the dependence of a response variable on one or more predictors, including prediction of future values of a response, discovering which predictors are important, and estimating the impact of changing a predictor or a treatment on the value of the response. It builds a model which defines the relationship between predictor and response variables. In order to solve the logistic regression, we introduced a stepwise feature selection method that considers both the dynamic and static characteristics of a program. The stepwise method considers static characteristic of the program statements while computing their grouping effects on the program termination state.

---

[1] The columns of ones in required for estimating $\beta_0$

In order to mitigate the instability problem of statistical analysis, the static features of the program constructs such as fault-proneness, inter-dependencies and reliability are also considered. In fact, while selecting the program statements into the regression model, we can consider the statements fault-proneness factors such as code complexity, debugging history, reliability, and the programmer's experiences. In this regard, the aim is to build a sparse model containing a very small number of program statements which highly affect the program result. To this end, we convert our problem into iteratively reweighted least squares or IRLS [41] equation using Newton-Raphson algorithm and then solve the equation using penalized least angle regression method [19]. The penalized technique we use to estimate the coefficients have several advantages. It has powerful feature selection capability to retain an appropriate number of fault relevant statements and eliminate the redundant and irrelevant ones. Due to applying both ridge and lasso penalties, the penalized method also handles situations when the number of program statements is much larger than the number of test cases. Other penalized techniques such as lasso and ridge regression do not have this significant capability.

The logistic regression method is used to learn functions of the form $f: X \rightarrow Y$ via the conditional probabilities of $Pr(Y|X)$ where $Y$ is a discrete value (binary in our application) and $X = <X_1, \dots X_n>$ is a vector of discrete or continuous variables (discrete non-zero in our application). Assuming a parametric form for the distribution $Pr(Y|X)$, logistic regression estimates the parameters (i.e. the weight or coefficient of each program statement in the model) from the SCM. The logistic regression model for the response binary variable $R = <r_1 \dots r_n>$ is in the conditional probability form:

$$Pr(R = 1|S = s) = \frac{1}{1 + e^{-(\beta_0 + s^T \beta)}} \tag{4}$$

and

$$Pr(R = 0|S = s) = \frac{1}{1 + e^{+(\beta_0 + s^T \beta)}} = 1 - Pr(R = 1|S = s) \tag{5}$$

where $\beta_0, \beta_1, \dots, \beta_n$ are the regression coefficients and s indicates the value of the spectrum S, extracted from a specific execution. Equations (4) and (5) specify the probability of program's execution to be passing and failing, respectively. These equations are called logistic functions. Knowing the number of times each statement is executed in a particular run $(S = s)$, equations (4) and (5) estimate the probability that the run is passing and failing, respectively. These functions are nonlinear in terms of $\beta_0$ to $\beta_n$. The logit transformation, converts a logistic function into a corresponding linear function as follows;

$$\log \frac{Pr(R = 1|s)}{Pr(R = 0|s)} = \beta_0 + s^T \beta \tag{6}$$

In logistic regression the coefficient values $\beta = <\beta_0, \beta_1, \dots, \beta_n>$ are chosen to maximize the conditional data likelihood which is the probability of the observed response conditioned on their corresponding $s$ values. We choose the parameters $\beta$ which satisfy:

$$\beta \leftarrow \arg\max_{\beta} \prod_i Pr(r_i|s_i, \beta) \tag{7}$$

To ease further calculations, we work on the maximum log-likelihood function by converting (7) to (8) as follows:

$$\log(\beta) \leftarrow \log\left(\arg\max_{\beta} \prod_i Pr(r_i|s_i, \beta)\right) \Rightarrow \beta \leftarrow \arg\max_{\beta} \sum_i \log Pr(r_i|s_i, \beta) \tag{8}$$

Expanding relation (8) by substituting $Pr(r_i|s_i,\beta)$ with right-hand side of the equation (4) and (5), the conditional log-likelihood $ll(\beta)$ is:

$$ll(\beta) \leftarrow \sum_{i=1}^{m} r_i \log Pr(y_i = 1|s_i,\beta) + (1 - r_i) \log Pr(y_i = 0|s_i,\beta) \tag{9}$$

Where $r_i$ indicates the program termination state value which is either 1 for passing or 0 for failing results. Equation (9) could be further simplified to:

$$ll(\beta) \leftarrow \sum_{i=1}^{m} r_i \beta^T s_i - \log(1 + e^{\beta^T s_i}) \tag{10}$$

To find the optimal values for $\beta$ the log-likelihood function in (10) is maximized by setting its derivatives to zero. The result is $n + 1$ score equations which are nonlinear in $\beta$. When the derivative of $ll(\beta)$ is zero, the corresponding data point, critical point, can either maximize or minimize the value of $ll(\beta)$. The critical point will be a maximum if the Hessian matrix of log-likelihood is nonpositive definite, which means $ll(\beta)$ is convex [42-43]. Since $ll(\beta)$ is convex, it has been shown in [42-43], the derivative to be zero will be necessary and sufficient condition for a data point to maximize the value of $ll(\beta)$.

$$\frac{\partial ll(\beta)}{\partial \beta} = \sum_{i=1}^{m} s_i(r_i - Pr(r_i = 1|s_i,\beta)) = 0 \tag{11}$$

Because the first element of $s_i$ is one, the first score equation can be written as follows:

$$\sum_{i=1}^{m} r_i = \sum_{i=1}^{m} Pr(r_i = 1|s_i,\beta) \tag{12}$$

To solve the score equations, different methods could be used. We apply the Newton-Raphson algorithm for which we need to know the Hessian matrix that is the second derivative of log-likelihood (Equation (11)):

$$\frac{\partial^2 ll(\beta)}{\partial \beta \partial \beta^T} = -\sum_{i=1}^{m} s_i s_i^T \, Pr(r_i = 1|s_i,\beta^{old})(1 - Pr(r_i = 1|s_i,\beta^{old})) \tag{13}$$

According to Newton-Raphson algorithm, the parameter estimation is done iteratively and in each iteration the new estimate, $\beta^{new}$, is computed by updating the old estimate, $\beta^{old}$:

$$\beta^{new} = \beta^{old} - \left(\frac{\partial^2 ll(\beta)}{\partial \beta \partial \beta^T}\right)^{-1} \frac{\partial ll(\beta)}{\partial \beta} \tag{14}$$

Here, the derivatives are evaluated at $\beta^{old}$. For more convenience, we rewrite the score equations and the Hessian in matrix notation. To this end, the following notations are used:

- $r = <r_1, r_2, \ldots, r_m>$ is the Boolean vector where $r_i = 1$ indicates passing and $r_i = 0$ denotes the failing termination of the $i^{th}$ execution of the subject program.
- $s: m * (n + 1)$ denotes the matrix of $s_i$ values.
- $P$ is the vector of fitted probabilities in which the $i^{th}$ component is: $Pr(r_i = 1|s_i,\beta^{old})$
- $W$ is a $m * m$ diagonal matrix of weights in which the $i^{th}$ diagonal element is: $Pr(r_i = 1|s_i,\beta^{old})(1 - Pr(r_i = 1|s_i,\beta^{old}))$

Using the above notations, equations (11) and (13) are rewritten as:

$$\frac{\partial ll(\beta)}{\partial \beta} = s^T(r - P) \tag{15}$$

$$\frac{\partial^2 ll(\beta)}{\partial \beta \partial \beta^T} = -s^T W s \tag{16}$$

Hence, the Newton step defined in (14) is rewritten as:

$$\beta^{new} = \beta^{old} + (s^T W s)^{-1} s^T (r - P) = (s^T W s)^{-1} s^T W \left( s\beta^{old} + W^{-1}(r - P) \right) = (s^T W s)^{-1} s^T W Z \tag{17}$$

As could be seen, the Newton step is expressed as a weighted least squares step in which the response is $Z$:

$$Z = s\beta^{old} + W^{-1}(r - P) \tag{18}$$

These equations are solved iteratively. In each iteration the value of $P$, $W$ and $Z$ are computed using their values in the previous iterations. This algorithm is referred to as *iteratively reweighted least squares* or *IRLS*, in which each iteration solves the known weighted least squares problem [35]:

$$\beta^{new} \leftarrow \underset{\beta}{\mathbf{argmin}}(Z - s\beta)^T W (Z - s\beta) \tag{19}$$

Typically, $\beta = \mathbf{0}$ is a good starting value for the iterative procedure. Let $W = \omega \omega^T$ where $\omega$ is a diagonal matrix using the Choleski Decomposition in which the $i^{th}$ component is $\sqrt{W_i}$. Thus the the above equation will be:

$$\beta^{new} \leftarrow \underset{\beta}{\mathbf{argmin}}(Z - s\beta)^T \omega^T \omega (Z - s\beta) = \underset{\beta}{\mathbf{argmin}}(\omega Z - \omega s\beta)^T (\omega Z - \omega s\beta) \tag{20}$$

Now assume that:

$$Y = \omega Z \text{ and } X = \omega s \tag{21}$$

Thus the above equation can be expressed as follows:

$$\beta^{new} \leftarrow \underset{\beta}{\mathbf{argmin}}(Y - X\beta)^T (Y - X\beta) \tag{22}$$

The above equation represents IRLS in a matrix notation format. In the previous section, we applied the Newton-Raphson method to convert nonlinear logistic equations, modeling a program behavior, into a corresponding linear least square problem, IRLS [41]. The ideal is to solve the IRLS problem, defined in equation (22), with a minimum number of bug predictor statements, $X$, while considering the static features and inter-dependencies of the statements. To consider the grouping effects of the statements, to be selected as bug predictors, on the program termination state the Elastic-Net constraints are imposed on the coefficients of the statements, to be selected as the regression variables. The Elastic-Net constraints are adjusted according to the FPL of the program statements. To achieve this, Lars method for stepwise solution of IRLS is employed.

### 2.4.1 Considering the static code fault-proneness during regression model construction

The experiments have shown that some parts of programs are more likely to contain faults. This fact should be considered while building the statistical fault localization model. In other words, all parts of a subject program should not be treated equally. The fault prone areas of code should be scored more cautiously. As mentioned above, the value of shrinkage parameter, highly affects the number of variables included in the final model. In equation (30), parameter $\xi$ controls the number of variables such that larger value of $\xi$ decreases the number of included variables and vice versa. Since, the program statements may have different static FPLs, we suggest using different values of $\xi$ for different groups of statements. To this end, in the case of fault-prone statements, we use $\xi$ with smaller values compared to not fault-prone statements. As a result, those statements are hardly eliminated from the corresponding model. Note that it does not mean we neglect the role of statements in less fault-prone locations on program failure. In the proposed approach, the program statements are categorized according to their static FPLs. For statements in fault-prone parts of the code, the value of the shrinkage parameter is chosen based on equation (23).

The default penalty value is 1 for each statement, but other values can be specified. In particular, any statement with penalty value equal to zero is not penalized at all. When penalizing the coefficients of statements, the value of their penalty parameters is determined according to their static FPL. Equation (23), presents how the penalty value is computed for a given statement $x$.

$$\begin{matrix} 1 - (FPL(x))^\theta & 3 \ll \theta \ll 4 & if\ FPL(x) \leq 0.5 \\ 1 - (FPL(x) * \theta) & 2 \ll \theta \ll 3 & if\ FPL(x) > 0.5 \end{matrix} \qquad (23)$$

For statements with static FPL smaller than 0.5, statements located in not fault-prone areas, penalty value will be close to 1. As a result, their shrinkage value will not be faced with any reduction. Penalty value will be reduced to 0.7 for statements located in the most fault-prone areas of the program. For other fault-prone statements, it will be in the range of (0.7-1).

### 2.4.2 Imposing Elastic-Net constraints

In order to provide grouping effect property for FPA-FL, which is helpful for finding multiple faults and correlated predictor variables, and to solve $P \gg N$ problem, we make use of the privileges of Elastic-Net method. To achieve this, we impose the following constraints on norm-1 and norm-2 of the regression coefficients:

$$L(\lambda_1, \lambda_2, \beta) = |y - X\beta|^2 + \lambda_2|\beta|^2 + \lambda_1|\beta|_1 \qquad (24)$$

Where $\lambda_1$ and $\lambda_2$ are the regularization parameters on norm-1 and norm-2 of the coefficients, respectively. Norm-2 and norm-1 are defined as follows;

$$|\beta|^2 = \sum_{j=1}^{n} \beta_j^2 \quad , \quad |\beta|_1 = \sum_{j=1}^{n} |\beta_j| \qquad (25)$$

According to the above equation, the estimated value for $\hat{\beta}$ minimizes the Elastic-Net:

$$\hat{\beta} = arg\ \min_{\beta}\{L(\lambda_1, \lambda_2, \beta)\} \qquad (26)$$

To convert the equation (24) to penalized least squares, we act as follows. Let $\alpha = \lambda_2/(\lambda_1 + \lambda_2)$; Hence, estimation of $\hat{\beta}$ is equal to the following optimization problem:

$$\hat{\beta} = arg\ \min_{\beta}|y - X\beta|^2 \text{ subject to } (1 - \alpha)\sum|\beta_i| + \alpha \sum \beta_i^2 \leq s_i \sum |\beta_i^{ols}| \qquad (27)$$

where $\beta_i^{ols}$ is the value of $\beta_i$ determined by ordinary least squares regression. The function $(1 - \alpha)|\beta|^1 + \alpha|\beta|^2$ is called elastic penalty which is a convex combination of lasso and ridge regression penalty parameters and $\alpha \in [0,1]$. $\alpha$ determines the balance between the lasso and ridge penalties, where $\alpha = 0$ amounts to lasso regression, and $\alpha = 1$ amounts to ridge regression. In practice, $\alpha$ is a vector, for each value of which we use 10-fold cross-validation to pick $s_i$. This approach amounts to a grid search of the parameter space as described in [23]. To incorporate prior information about program statements directly into the model selection approach, we minimize Equation (27) subject to a new penalty function:

$$(1 - \alpha) \sum |\delta_i \beta_i| + \alpha \sum \beta_i^2 \leq s_i \sum |\beta_i^{ols}| \qquad (28)$$

where $\delta_i$ is a modifier on the shrinkage incurred on each statement. If there is a prior belief for a causal relationship $x_p \rightarrow y_p$, then $\delta_i < 1$ corresponds to less shrinkage being incurred on the corresponding $\beta_i$, hence making it more likely that this statement is not shrunk out of the model. Note that neither the correlation with a target nor the order in which predictors selected by the model are modified and only the degree of shrinkage of a parameter is altered. In cases where multiple predictors are correlated (a common occurrence in programs), $\delta_i$ will cause predictors with no prior information to be shrunk from the model before predictors with prior information. Note that the $\delta_i$ modifies only the $l_1$ norm, as in [44]. This implementation is based on the elasticnet R package [44]. Given dataset $(y, X)$ and $(\lambda_1, \lambda_2)$, define an artificial dataset $(y^*, X^*)$ by

$$X^*_{(n+p) \times p} = \frac{1}{\sqrt{1+\lambda_2}} \begin{bmatrix} X \\ \sqrt{\lambda_2} I \end{bmatrix} \quad \text{and} \quad y^*_{(n+p)} = \begin{bmatrix} y \\ 0 \end{bmatrix} \quad (29)$$

Let $\xi = \lambda_1 / \sqrt{1+\lambda_2}$ and $\beta^* = \sqrt{1+\lambda_2}\beta$. Then, the Elastic-Net estimator is equivalent to the following lasso estimator:

$$\hat{\beta} = \sqrt{1+\lambda_2}\hat{\beta}^* \text{ where } \hat{\beta}^* = arg \min_{\beta^*}(\|y^* - X^*\beta^*\|^2 + \xi\|\beta\|_1^*) \quad (30)$$

By transforming Elastic-Net equation into lasso with embedded ridge penalty into the matrix, $P >> N$ problem could be handled. In other words, when the number of program statements is greater that the number of test cases, the possibility of solving the equation will be provided using the augmented matrix.

### 2.4.3 Combining the results of static fault-proneness analysis and dynamic Elastic-Net

Four possible scenarios should be considered while incorporating the static FPLs into statistical fault localization:

1- Given that statement $x$ is located in a fault-prone area of the program and it is assigned a high suspiciousness score by the Elastic-Net method. In this case, since $x$ is recognized as fault-prone by static analysis, it will gain a high static FPL. On the other hand, since $x$ is assigned a high suspiciousness score by the Elastic-Net method, it should not be appeared in many passing executions. Consequently, it is expected that static FPL of $x$ is not diminished significantly during the refinement process. In this scenario, the maximum of the two scores obtained by the dynamic and static analysis methods is considered as the final score.

2- Given that statement $x$ is located in a not fault-prone (NFP) area of the program and it is assigned a high suspiciousness score by the Elastic-Net method. Since $x$ is not recognized as fault-prone by static analysis, it will gain a low static FPL. In this case, we should not let low static FPL of $x$ causes a reduction in its final suspiciousness score. Because it is not necessary for all the program faults to be found around the complex and fault-prone areas. In this case, we have considered the suspiciousness score obtained by the Elastic-Net method as the final score.

3- Given that statement $x$ is located in fault-prone (FP) area of the program and it is assigned a low suspiciousness score by the Elastic-Net method. In this case, since $x$ is recognized as fault-prone by static analysis, it will gain a high static FPL. During the refinement process three possible cases may occur: 1) Since the statement $x$ has gained a low suspiciousness score by the Elastic-Net method, it may appears in very passing executions and has high correlation with passing state of the program rather than failing. In this case, the static FPL of $x$ will be reduced significantly during the refinement process and it will gain a low score in the final result. 2) Suspiciousness score of $x$ estimated by the Elastic-Net method is low because $x$ has not appeared in many of passing and failing executions. In this case, $x$ may not be considered as faulty statement and prevented to enter the model. This situation may be occurred due to inappropriateness of existing test suite. In this case, provided that the static FPL of $x$ is not reduced significantly during the refinement process, taking into account the static FPL of $x$ may be helpful and the simple average of two scores can be considered as the final score. 3) Suspiciousness score of $x$ estimated by the Elastic-Net method is low because our method may be unable to localize all types of faults. In this case, similar to second situation, taking into consideration the static FPL of $x$ may be helpful and the simple average of two computed scores can be considered as the final score.

4- Given that statement $x$ is located in not fault-prone (NFP) area of the program and it is assigned a low suspiciousness score by the Elastic-Net method. In this case, since $x$ is recognized as not fault-prone by static analysis, it will gain a low static FPL. In this case, we have considered the suspiciousness score obtained by the Elastic-Net method as the final score.

Elastic-Net can encourage a grouping effect, i.e., strongly correlated statements tend to be selected in groups. Since fault relevant statements may be highly correlated with a few fault irrelevant statements, the Elastic-Net would include redundant statements into the fitted model. Hence, we need to further investigate the fault relevant statements within the selected groups. Proposed approach of combining the results are able to alleviate this problem to some extent. In fact, the fault irrelevant statements within each group will be faced with weight loss either they have been located in fault-prone areas or not. It is because the redundant statements are included in a group just because of existing high correlation among grouped statements. Since redundant statements are not highly correlated with failure and are not fault relevant, the likelihood of observing these statements in passing executions is more than in failing ones. So, refinement process will reduce the static FPLs of redundant statements included in significant groups. A significant group is a group of statements which are highly correlated with each other and contains fault relevant statements. Consequently, we will be capable of ranking statements within a group based on static FPLs of included statements.

Since the range of the output values differs for two methods normalization of data in the range of [0,1] is required before combining the results. We normalize suspiciousness scores from 0 to 1 (both inclusive) for dynamic method suspiciousness rankings. That is, the normalized suspiciousness is defined as:

$$Norm_{Susp} = |susp| - |min|/|max| - |min| \tag{31}$$

Where $min$ and $max$ denote the minimum and the maximum scores obtained by the Elastic-Net method. The scores of two methods are further synthesized to form the final ranking. Later the output is sorted in decreasing order and can be used to construct cause-effect chains.

### 2.4.4 Constructing Cause-Effect Chains

After identification of fault-relevant statements using the proposed hybrid method, we need to discover how these statements relate to each other in program dependency graph. Determining cause-effect chains, dependency chains that connect faulty and fault-relevant statements in static program dependency graph, allows a programmer to focus his attention on the most failure relevant parts of programs. As stated before, most of program faults are complex and involve complex interactions between many program statements. Such interactions among statements can provide additional information about faults. For example, we can exploit the transitional information provided by data/control dependencies among program statements to expose more bugs and discover cause-effect chains. In this regard, the grouping effect of the Elastic-Net method could be very helpful. Due to grouping effect of the Elastic-Net method, all fault correlated statements are retained in the model and taken similar coefficients. Elastic-Net puts highly correlated statements into a single group. Therefore, the fault relevant statements corresponding to each individual bug are included into a group separated from groups including faulty statements of other existing bugs of a program. So, this property is helpful for constructing cause-effect chain(s) for statements included in a group and to find multiple faults in programs.

For each group of statements identified by the Elastic-Net method, we analyze the static dependency graph of the original program and find those dependency chains that connect all the statements. Since identified chains are often not unique, we select the shortest one as the cause-effect chain for manual inspection.

## 3. Experimental Evaluation

In this section, the effectiveness of FPA-FL is empirically evaluated. To this end, we compare the proposed method with some well-known SBFL techniques in the context of software fault localization. To show the performance of FPL-FL, the following case studies are designed to evaluate the proposed method in different ways:

1- The performance of FPL-FL in finding the origin of failures is measured according to various evaluation frameworks and compared to the state-of-the-art SBFL techniques. We evaluate FPA-FL on seven programs of the Siemens test suite, which is utilized by many researchers to evaluate their methods [24][45-46], moderately larger programs including Gzip, Grep, Sed, and Space, and Defects4J programs. We also investigate the scalability of proposed method by evaluating FPA-FL on Make and bash programs which are considered as large programs.
2- In order to investigate how grouping effect of the Elastic-Net method could give the capability of finding multiple bugs to FPA-FL, we combine several versions of subject programs to generate multiple-bug versions. After conducting experiments, results are compared to fault localization techniques that are able to localize multiple faults in programs.
3- The impact of considering static fault-proneness analysis and slice coverage spectrum on fault-localization is studied in case study 3. We conducted experiments on the subject programs both considering and not considering the static fault-proneness analysis and reported results. We also evaluate the effectiveness of proposed slice coverage spectrum which is based on EBDSs of program executions.

All experiments in this section are carried out on a 2.66 GHz Intel Core 2 Quad Processor PC with 6 GB RAM, running UBUNTU 9.10 Desktop (i386). To perform clustering, we applied WEKA machine learning and data mining tool [47] which is open source software that implements clustering algorithms, accurately. Static complexity code metrics are calculated by imagix4D [48] to perform the fault-proneness analysis. Elastic-Net regression model is constructed using the *glmnet* package [44].

### 3.1 Subject Programs

We use the Siemens suite, Defects4J suite, Gzip, Grep, Sed, Space, Make and Bash as our subject programs to evaluate FPA-FL. The Siemens suite contains seven subject programs with 132 faulty versions and each faulty version contains one manually seeded fault. As stated in [49], a fault localization technique is valuable if it works on real faults. According to [50], artificial faults differ from real faults in many respects, including their size, their distribution in code, and their difficulty of being detected by tests. So, we also used some programs of the Defects4J [51] suite (v1.1.0), which consists of 224 real faults from 4 open source projects: JFreeChart, Commons Lang, Commons Math, and Joda-Time. For each real fault, Defects4J provides a developer-written test suite containing at least one such fault-triggering test case. Version 1.1.2 of the Gzip program was downloaded from Software Infrastructure Repository (SIR). Also found at SIR were versions 1.2 of the Grep, 3.76.1 of the make, 2.0 of the Sed, 2.0 of the Space and 1.0 of the Bash programs. A brief description of the test suites is presented in Table 1.

Table 1. A brief description of test suites which are used to evaluate the performance of FPA-FL

| Program | Faulty versions(Seeded/Real Faults) | # of Lines | # of Test cases | Language |
|---|---|---|---|---|
| Siemens suite | | | | |
| Print tokens | 7- all seeded faults | 472 | 4056 | C |
| Print tokens2 | 10- all seeded faults | 399 | 4071 | C |
| Replace | 32- all seeded faults | 512 | 5542 | C |
| Schedule | 9- all seeded faults | 292 | 2650 | C |
| Schedule2 | 10- all seeded faults | 301 | 2680 | C |
| Tcas | 41- all seeded faults | 141 | 1578 | C |
| Tot info | 23- all seeded faults | 440 | 1054 | C |
| Defects4J | | | | |
| JFreeChart | 26- all real faults | 96K | 2,205 | Java |
| Joda-Time | 27- all real faults | 28K | 4,130 | Java |
| Commons Lang | 65- all real faults | 22K | 2,245 | Java |
| Commons Math | 106- all real faults | 85K | 3,602 | Java |
| Other programs | | | | |
| Gzip | 55- all seeded faults | 6K | 217 | C |
| Grep | 17- all seeded faults | 12K | 809 | C |
| Sed | 17-real and seeded faults | 12K | 370 | C |
| Make | 31- all seeded faults | 20K | 793 | C |
| Space | 38- real and seeded | 9K | 13585 | C |
| Bash | 6- all seeded faults | 59K | 1168 | C |

## 3.2 Evaluation metrics

Performance metrics are widely used to facilitate comparisons among different approaches. Some evaluation metrics has been used by state-of-the-art approaches like Ochiai [4], D-Star [5], O-O$^P$ [6], H3b-H3c [7] and Baah [52-53] methods. Two metrics are used in this paper:

1) '*The EXAM score*' gives the percentage of statements that need to be examined until the first faulty statement is reached. According to [5] the main objective is to provide a good starting point for programmers to begin fixing a bug, rather than identifying all the code that would need to be corrected. From this perspective, even though a fault may span multiple, non-contiguous statements, the fault localization process is halted once the first statement corresponding to the bug is found.

2) '*Average number of statements examined*' metric gives the average number of statements that need to be examined with respect to a faulty version (of a subject program) to find the bug. If a program has n faulty versions, we can say, for example, A is more effective than B, if $\frac{\sum_{i=1}^{n} A(i)}{n} < \frac{\sum_{i=1}^{n} B(i)}{n}$ where $A(i)$ and $B(i)$ are the number of statements need be examined to locate the fault in the $i$*th* faulty version by A and B, respectively.

We note that, for *FPA-FL* and all other peer techniques discussed herein, the same suspiciousness value may be assigned to multiple statements, and therefore be tied for the same position in the ranking. So, the results are provided in two different levels of effectiveness – the ''best'' and the ''worst''. In all our experiments we assume that for the ''best'' effectiveness we examine the faulty statement first and for the ''worst'' effectiveness we examine the faulty statement last.

## 3.3 Case study 1: finding the location of single bugs

In this study, we evaluate the performance of FPA-FL in finding the location of single bugs and compare it to the state-of-the-art fault localization techniques using the *EXAM* and *Average Number of Statements Examined* metrics. We subject FPA-FL and other peer techniques to the various faulty programs and measure their localization quality. Eight distinguished fault localization techniques include Baah [52-53], Ochiai [4], D-Star [5], O-O$^P$ [6] and H3b-H3c [7] are compared to FPA-FL in this experiment.

Table 2  Average number of statements examined with respect to each faulty version (Best case)

| Technique | Siemens | Gzip | Grep | Sed | Space | Bash | Make | J-Chart | J-Time | C-Lang | C-Math |
|---|---|---|---|---|---|---|---|---|---|---|---|
| FPA-FL | **9.21** | **51.2** | **102.94** | **66.32** | **35.9** | 76.33 | **115.84** | **4120.18** | **177.62** | **1238.48** | **2782.54** |
| D-Star | 14.66 | 58.82 | 148.21 | 79.38 | 43.28 | 92.46 | 271.62 | 4821.72 | 239.52 | 1408.65 | 3021.25 |
| H3b | 13.21 | 76.48 | 165.06 | 125.25 | 59.10 | 128.50 | 304.39 | 4872.15 | 235.42 | 1421.85 | 2956.11 |
| H3c | 13.21 | 76.48 | 145.63 | 122.10 | 57.95 | 122.63 | 246.79 | 4872.15 | 229.67 | 1411.56 | 2916.69 |
| Ochiai | 15.57 | 68.10 | 164.50 | 81.45 | 44.25 | **74.67** | 275.11 | 4770.90 | 241.49 | 1449.78 | 3046.19 |
| O | 10.21 | 56.78 | 125.06 | 68.45 | 37.12 | 91.89 | 189.36 | 5359.24 | 342.51 | 1479.94 | 4148.72 |
| O$^P$ | 11.02 | 56.78 | 125.06 | 68.45 | 38.50 | 91.89 | 189.36 | 5448.62 | 362.14 | 1499.48 | 4182.42 |
| Baah, 2010 | 16.55 | 78.48 | 195.45 | 145.25 | 52.31 | 96.63 | 152.25 | 4764.72 | 254.38 | 1412.26 | 2958.72 |
| Baah, 2011 | 12.82 | 61.48 | 155.63 | 118.10 | 45.85 | 85.44 | 141.85 | 4625.38 | 224.50 | 1389.92 | 2914.65 |

Table 3  Average number of statements examined with respect to each faulty version (Worst case)

| Technique | Siemens | Gzip | Grep | Sed | Space | Bash | Make | J-Chart | J-Time | C-Lang | C-Math |
|---|---|---|---|---|---|---|---|---|---|---|---|
| FPA-FL | **18.85** | **119.52** | **191.45** | 176.26 | 63.41 | **106.75** | 248.36 | **4851.55** | 246.47 | **1424.28** | 3562.80 |
| D-Star | 22.69 | 126.59 | 225.16 | 181.51 | 82.81 | 127.62 | 452.78 | 5658.26 | 294.52 | 1682.64 | 3945.27 |
| H3b | 19.80 | 147.23 | 255.63 | 231.85 | 95.08 | 206.82 | 483.32 | 5708.82 | 299.26 | 1698.11 | 3879.36 |
| H3c | 19.46 | 147.23 | 236.19 | 228.70 | 92.56 | 195.45 | 425.71 | 5108.82 | 281.75 | 1684.56 | 3824.12 |
| Ochiai | 23.62 | 136.90 | 255.06 | 198.05 | 74.89 | 136.92 | 454.04 | 5682.46 | 351.49 | 1699.78 | 4026.19 |
| O | 31.72 | 131.52 | 215.63 | **173.05** | **62.89** | 116.81 | 368.29 | 5848.75 | 392.69 | 1714.85 | 4841.56 |
| O$^P$ | 22.29 | 131.52 | 215.63 | **173.05** | 64.34 | 116.81 | 368.29 | 5969.26 | 446.74 | 1736.62 | 4926.48 |
| Baah, 2010 | 29.80 | 169.33 | 275.43 | 231.85 | 69.80 | 169.55 | 312.96 | 5548.83 | 344.28 | 1605.42 | 3916.92 |
| Baah, 2011 | 23.46 | 139.29 | 236.19 | 210.70 | 68.46 | 140.25 | 295.38 | 5359.27 | 329.65 | 1582.39 | 3892.64 |

Tables 2 and 3 present the average number of statements that need to be examined by each fault localization technique across each of the subject programs, for both best and worst cases. For example, the average number of statements examined by *FPA-FL* with respect to the all faulty versions of Gzip is 51.2 in the best case and 119.52 in the worst. Again, with respect to the Gzip program, we find that the second best technique is O, which can locate all the faults by requiring the examination of no more than 56.78 statements in the best case, and 131.52 in the worst. For D-Star, the best is 58.82 and the worst 126.59. A similar observation also applies to other programs. In the case of Space, *FPA-FL*, in the worst case, detects the faults by examining 63.41 statements in average whereas O and O$^P$ do so in 62.89 and 64.34 statements, respectively.

To investigate the impact of $\theta$, in static FPL refinement process, on the effectiveness of FPA-FL, we provide the average number of statements examined by FPA-FL using different $\theta$'s values on all subject programs. It is observed that the best results are obtained with $\theta = 1.75$.

Table 4  Average number of statements examined by FPA-FL using different $\theta$'s values on all subject programs

|  | $\theta = 1.25$ | $\theta = 1.5$ | $\theta = 1.75$ | $\theta = 2$ |
|---|---|---|---|---|
| FPA-FL -Best | 811.52 | 799.21 | **797.87** | 808.62 |
| FPA-FL -Worst | 1036.80 | 1006.42 | **1001.79** | 1020.75 |

We now present the evaluation of FPA-FL with respect to the EXAM score. Figure 2 illustrates the EXAM score of FPA-FL and other peer techniques on subject programs. The x-axis represents the percentage of code (statements) examined while the y-axis represents the number of faulty versions where faults are located by the examination of an amount of code less than or equal to the corresponding value on the x-axis. For example, referring to Part (a1) of Figure 2, we find that by examining 10 % of the code *FPA-FL* can locate 81% of the faults in the Siemens suite in the best cases and 64% in the worst, whereas D-Star has 64% (best) and 52% (worst).

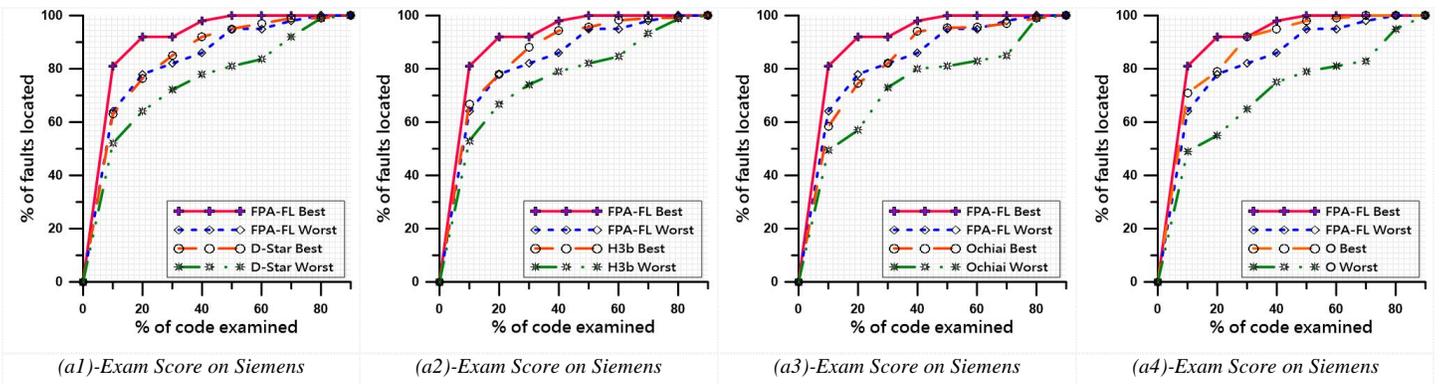

*(a1)-Exam Score on Siemens*   *(a2)-Exam Score on Siemens*   *(a3)-Exam Score on Siemens*   *(a4)-Exam Score on Siemens*

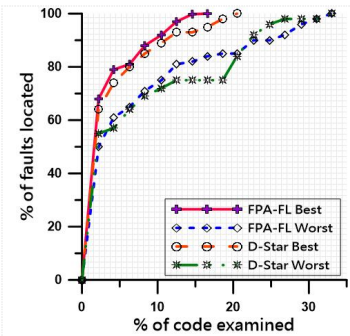
(b1) -*Exam* Score on Gzip

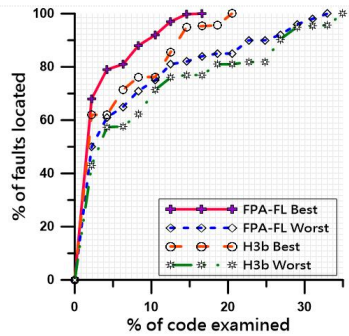
(b2) -*Exam* Score on Gzip

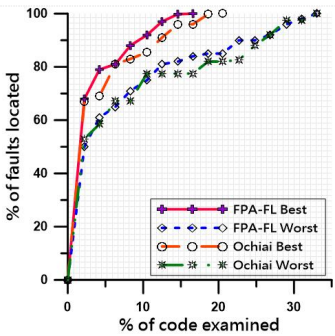
(b3) -*Exam* Score on Gzip

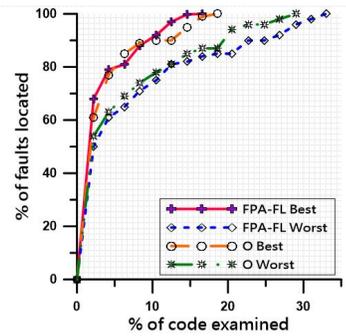
(b4) -*Exam* Score on Gzip

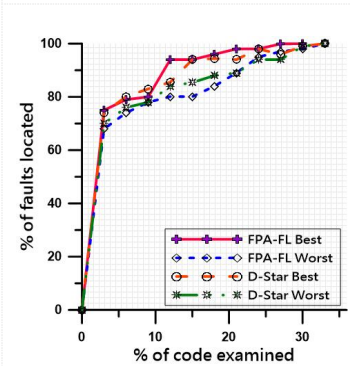
(c1) -*Exam* Score on Sed

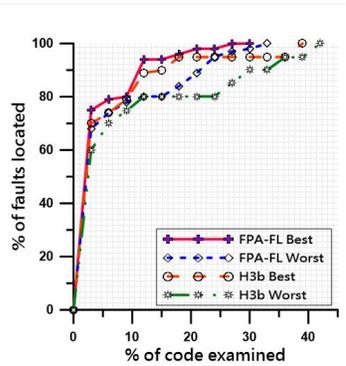
(c2) -*Exam* Score on Sed

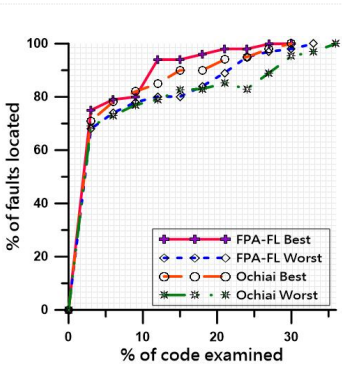
(c3) -*Exam* Score on Sed

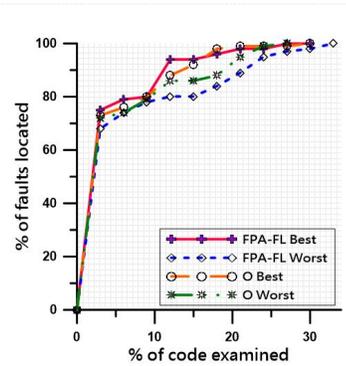
(c4) -*Exam* Score on Sed

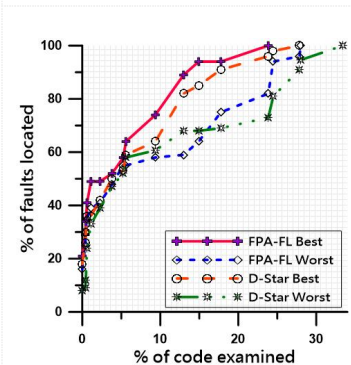
(d1) -*Exam* Score on Grep

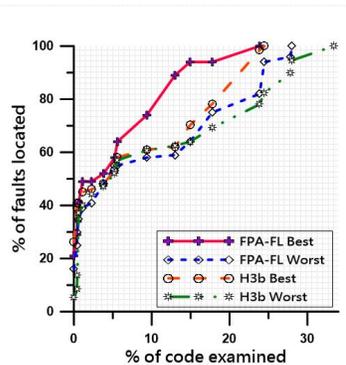
(d2) -*Exam* Score on Grep

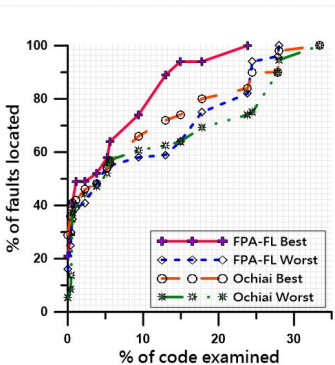
(d3) -*Exam* Score on Grep

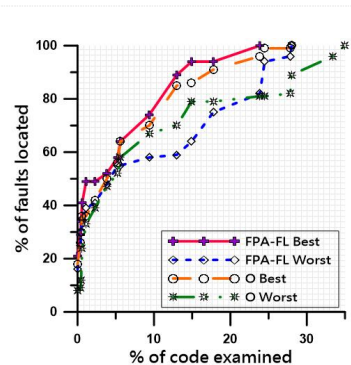
(d4) -*Exam* Score on Grep

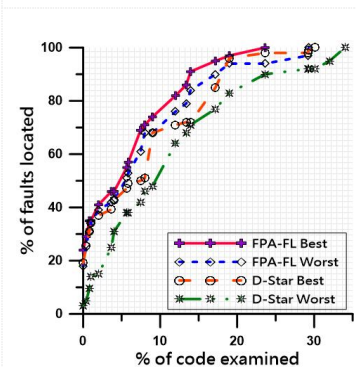
(e1) -*Exam* Score on Make

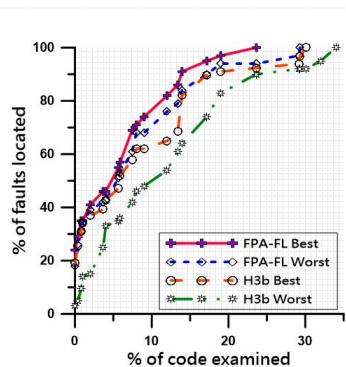
(e2) -*Exam* Score on Make

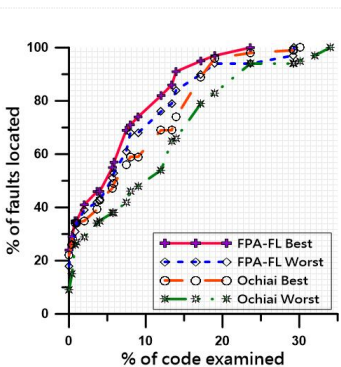
(e3) -*Exam* Score on Make

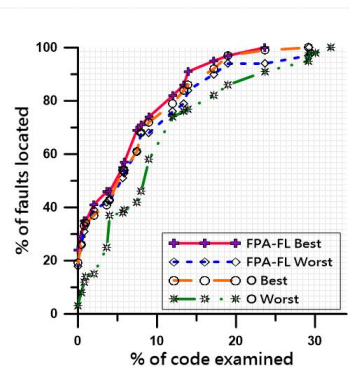
(e4) -*Exam* Score on Make

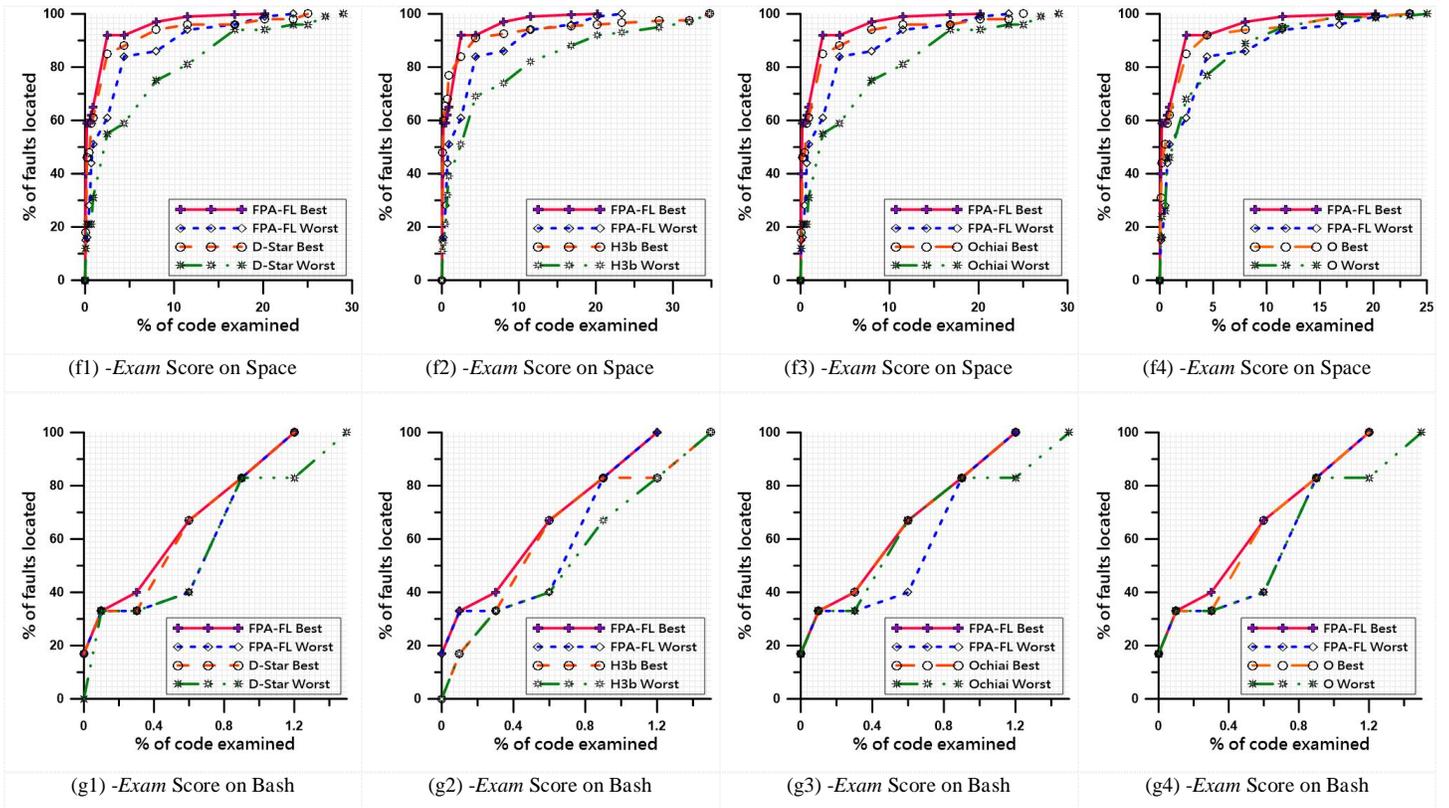

Figure 2. *EXAM* score-based comparison on subject programs

Many fault localization approaches [2-8][20] are evaluated on artificial bugs. However, it is unclear whether such bugs capture true characteristics of real bugs in real programs. Therefore, we also evaluate FPA-FL and other techniques on real bugs from 4 different software projects in the Defects4J suite [51], a database of real, isolated, reproducible software faults from real-world open-source Java projects intended to support controlled studies in software testing. The common way for evaluating fault localization handles defects that consist of a change to one executable statement in the program, as is the case for seeded faults. However, as reported in [51], most of real-world bug fixes span multiple statements. In our experiments, we consider that localization of any one defective statement is sufficient to understand and repair the defect.

Figure 3 illustrates the average EXAM score of FPA-FL and other peer techniques on Defects4J programs. It is observed that FPA-FL outperforms other techniques in the case of real faults. In fact, a large number of faults in large-sized real world programs are complex (i.e., they involve complex interactions amongst many program statements). Therefore, modeling the combinatorial effect of statements on each other and on the program termination state, could considerably improve fault localization. We also observed that the interaction between faulty statement(s) and other correlated statements are likely to cause coincidental correctness. To reduce the negative impact of the coincidentally correct tests on fault localization performance and to be able to localize this type of bugs, it is required to analyze the joint impact of the statements on the program failure. As stated before, the Elastic-Net regression method can be best applied to model the combinatorial effect of statements, because it considers the simultaneous effect of statements on each other as well as their effect on the program termination state.

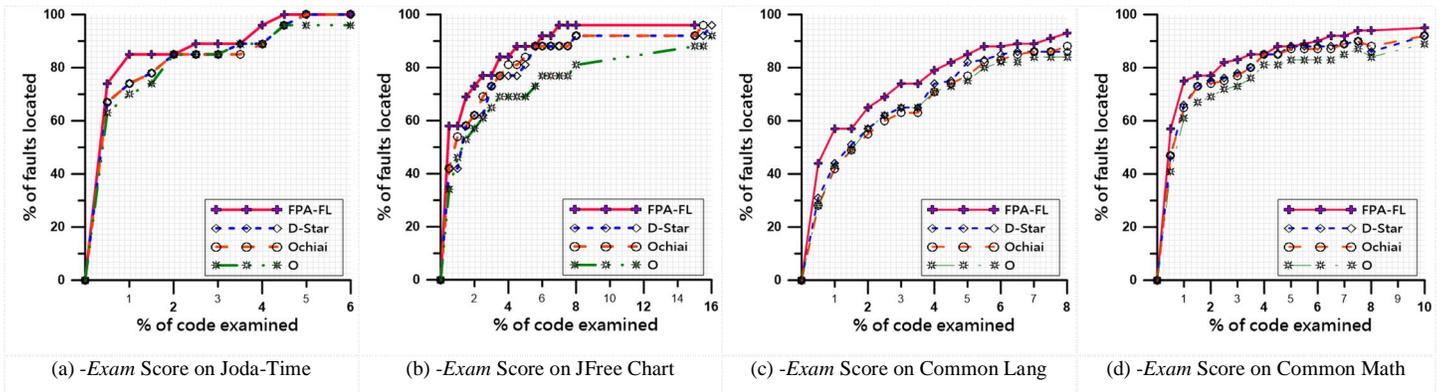

| (a) -*Exam* Score on Joda-Time | (b) -*Exam* Score on JFree Chart | (c) -*Exam* Score on Common Lang | (d) -*Exam* Score on Common Math |

Figure 3. *EXAM* score-based comparison on subject programs containing real faults

To show the effectiveness of FPA-FL on larger programs and how to deal with P>>N problem, we have also conducted our experiments on the Bash program with 59,846 lines of code. To demonstrate the ability of FPA-FL in constructing a stable model, four Elastic-Net regression models are constructed using 100, 200, 500 and 1000 test cases randomly selected from test pool. The number of test cases in this experiment is considerably less than the number of examined program statements, which is about 4200. Results are shown in Figure 4 as we can see, despite the small number of test cases, particularly in the subsets of 100 and 200, FPA-FL can found a significant percent of faults with a small amount of manual code inspection.

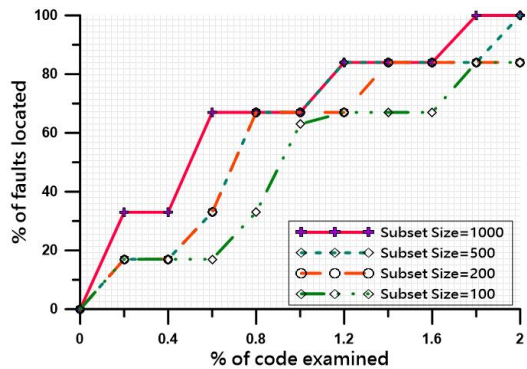

Figure 4. Performance of FPA-FL on Bash program with limited number of test cases

## 3.4  Case study 2: finding the location of multiple bugs

Programs that contain multiple faults present new challenges for fault localization. Most published fault-localization techniques target the problem of localizing a single fault in a program that contains only a single fault. In fact, the single-bug algorithms are not efficient for finding multiple bugs in programs. Previous techniques generally deal with multiple faults by the creation and specialization of test suites that target different faults. For example, automatic partitioning of the failing tests is widely used to locate multiple faults in programs [22]. This technique attempts to create subsets of the original test suite that target individual faults. However, in [54] it has been shown that the high degree of failures caused by multiple faults suggest that subsets are unlikely to be pure and target an individual fault.

In this case study, the effectiveness of FPA-FL is evaluated on programs with multiple bugs. A multiple fault version is a faulty version $X$ with $k$ faults that is made by combining $k$ faulty versions from a set $\{x_1; x_2; \ldots; x_k\}$ where each bug $i$ in $X$ corresponds to the faulty version $x_i$. In practice, developers are aware of the number of failing test cases for their programs, but are unaware of whether a single fault or many faults caused those failures. Thus, developers usually target one fault at a time in their debugging.

In the first experiment we randomly combined some faulty versions of subject programs to make multiple-bug versions (100 multiple-fault versions for each program) and then investigated that in what percent of the programs, all faults can be localized regarding to one, five, ten and twenty percent of manual code inspection. The result of experiments is shown in Figure 5.

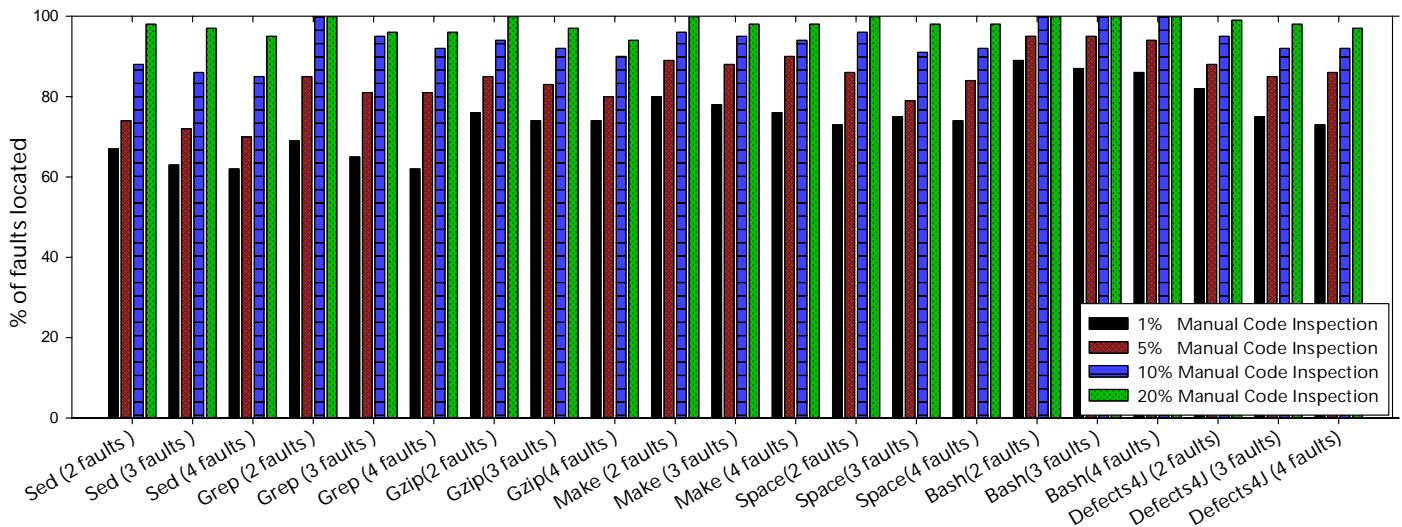

Figure 5: Quality Comparison in multiple faults

For programs with multiple faults, the authors of [55] also define an evaluation metric, *Expense*, corresponding to the percentage of code that must be examined to locate the first fault as they argue that this is the fault that programmers will begin to fix. We note that the *Expense* score, though defined in a multi-fault setting, is the same as the *EXAM* score used in this paper. The *Expense* score is really part of a bigger process that involves locating and fixing all faults (that result in at least one test case failure) that reside in the subject program. After the first fault has successfully been located and fixed, the next step is to re-run test cases to detect subsequent failures, whereupon the next fault is located and fixed. The process continues until failures are no longer observed, and we conclude (but are not guaranteed) that there are no more faults present in the program. This process is referred to as the *one-fault-at-a-time* approach, and thus the *Expense* score only assesses the fault localization effectiveness with respect to the first iteration of the process.

Since there is more than one way to create a multi-bug version, using only one may lead to a biased conclusion. To overcome this problem, 30 distinct faulty versions with 3, 4, 5 and 6 bugs, respectively, for Gzip, Grep, Sed, Space, Make, Bash and Defects4J programs are created. Also, a total of 120 multi-fault programs are created based on combinations of the single-fault programs of the Siemens suite, and they range from faulty versions with 2 faults to those with 5 faults.

Data in Table 5 gives the total number of statements that need to be examined to find the first fault across all 120 faulty versions of Siemens suite. We observe that, regardless of best or worst case, FPA-FL is the most effective.

Table 5  Total number of statements examined for the 120 multi-fault versions of the Siemens suite by FPA-FL and other techniques

| Technique | Best Case | Worst Case | Technique | Best Case | Worst Case |
|---|---|---|---|---|---|
| FPA-FL | **1274** | **2108** | Ochiai | 1561 | 2216 |
| D-Star | 1491 | 2198 | O | 1438 | 3951 |
| H3b | 4620 | 5424 | $O^p$ | 1899 | 2455 |
| H3c | 3492 | 4264 | Baah, 2011 | 1612 | 2356 |

We then apply *FPA-FL* and other representative techniques to faulty versions with multiple bugs using *one-bug-at-a-time* method. Tables 6 and 7 give the average number of statements that need to be examined to find the first bug for the best and the worst cases. For example, the average number of statements examined by *FPA-FL* for the five-bug version of Gzip is 38.51 for the best case and 89.59 for the worst, whereas 138.75 (best) and 215.48 (worst) statements needs to be examined by Ochiai and 119.62 (best) and 182.5 (worst) by D-Star.

Table 6  Average number of statements examined to locate the first bug (best case)

| Technique | Gzip | | | | Grep | | | | Sed | | | |
|---|---|---|---|---|---|---|---|---|---|---|---|---|
| | 3-bug | 4-bug | 5-bug | 6-bug | 3-bug | 4-bug | 5-bug | 6-bug | 3-bug | 4-bug | 5-bug | 6-bug |
| FPA-FL | **39.41** | **44.26** | **38.51** | **46.66** | 141.25 | **171.32** | **215.16** | 119.26 | **50.51** | 64.42 | 53.69 | 61.20 |
| D-Star | 101.68 | 98.51 | 119.62 | 126.41 | **135.66** | 219.56 | 281.82 | 112.43 | 62.51 | 69.85 | 172.66 | 195.41 |
| H3b | 71.52 | 69.21 | 88.6 | 62.7 | 211.65 | 277.48 | 289.55 | **80.26** | 188.21 | 246.59 | 43.30 | 59.52 |
| H3c | 79.26 | 78.85 | 88.1 | 69.42 | 231.46 | 285.04 | 271.00 | 79.63 | 181.51 | 209.17 | **45.26** | **52.18** |
| Ochiai | 105.88 | 112.56 | 138.75 | 151.88 | 189.76 | 258.56 | 281.73 | 126.11 | 51.65 | **59.45** | 276.96 | 355.6 |
| O | 55.12 | 53.89 | 53.58 | 51.65 | 289.65 | 276.5 | 272.11 | 176.98 | 58.45 | 61.82 | 61.26 | 58.41 |
| O<sup>P</sup> | 169.54 | 158.8 | 151.45 | 151.23 | 289.65 | 276.5 | 272.11 | 176.98 | 218.51 | 319.79 | 376.52 | 251.63 |
| Baah, 2011 | 79.20 | 126.55 | 141.65 | 149.31 | 194.28 | 270.56 | 236.50 | 117.76 | 74.56 | 62.55 | 166.82 | 146.24 |

Table 6  continued

| Technique | Make | | | | Space | | | | Bash | | | |
|---|---|---|---|---|---|---|---|---|---|---|---|---|
| | 3-bug | 4-bug | 5-bug | 6-bug | 3-bug | 4-bug | 5-bug | 6-bug | 3-bug | 4-bug | 5-bug | 6-bug |
| FPA-FL | **124.89** | **156.56** | 177.33 | **50.33** | **98.23** | **84.95** | 97.41 | **78.32** | **84.64** | 159.95 | 122.37 | 98.90 |
| D-Star | 201.52 | 185.36 | **151.21** | 64.25 | 121.85 | 142.85 | 110.52 | 101.81 | 139.52 | 188.60 | 149.66 | 129.82 |
| H3b | 381.8 | 360.50 | 358.45 | 106.12 | 192.52 | 178.58 | 198.45 | 166.11 | 278.45 | 251.84 | 355.21 | 192.49 |
| H3c | 460.12 | 491.56 | 521.6 | 146.11 | 216.46 | 236.40 | 271.52 | 156.88 | 351.66 | 308.26 | 385.49 | 221.16 |
| Ochiai | 211.28 | 199.53 | 156.89 | 65.21 | 136.59 | 119.26 | 105.17 | 88.63 | 171.60 | 241.74 | 204.25 | 142.45 |
| O | 206.50 | 250.59 | 278.11 | 108.46 | 195.85 | 209.82 | 221.74 | 159.66 | 185.30 | 220.15 | 253.36 | 136.20 |
| O<sup>P</sup> | 369.21 | 482.74 | 499.25 | 151.63 | 221.62 | 251.45 | 298.25 | 172.45 | 166.82 | 266.36 | 281.81 | 159.24 |
| Baah, 2011 | 236.7 | 184.35 | 181.85 | 102.45 | 115.26 | 107.66 | **96.44** | 102.82 | 144.25 | 175.54 | 159.68 | 126.58 |

Table 6  continued

| Technique | J-Chart | | | J-Time | | | C-Lang | | | C-Math | | |
|---|---|---|---|---|---|---|---|---|---|---|---|---|
| | 3-bug | 4-bug | 5-bug | 3-bug | 4-bug | 5-bug | 3-bug | 4-bug | 5-bug | 3-bug | 4-bug | 5-bug |
| FPA-FL | **3863.2** | **2350.4** | **4150.7** | **198.2** | **246.8** | **201.8** | **892.6** | **1394.4** | **1760.4** | **2063.1** | **3261.2** | **2451.8** |
| D-Star | 4326.6 | 3259.2 | 4994.5 | 284.4 | 269.6 | 321.4 | 1268.5 | 1682.6 | 1896.9 | 3262.8 | 3757.4 | 2965.3 |
| H3b | 3952.8 | 3562.9 | 4324.1 | 325.1 | 254.7 | 345.2 | 1322.5 | 1752.2 | 1652.6 | 3425.7 | 3425.6 | 3016.3 |
| H3c | 4025.4 | 3644.3 | 4561.5 | 334.8 | 269.5 | 355.4 | 1352.4 | 1885.8 | 1826.6 | 3521.3 | 3642.8 | 3152.4 |
| Ochiai | 4152.5 | 3026.2 | 4521.4 | 261.2 | 265.6 | 305.3 | 1214.3 | 1425.5 | 1794.5 | 2895.6 | 3621.5 | 2755.2 |
| O | 4562.3 | 3366.1 | 4560.8 | 341.6 | 328.7 | 342.8 | 1358.1 | 1852.4 | 1946.6 | 3521.5 | 3952.4 | 3154.6 |
| O<sup>P</sup> | 4125.4 | 2852.4 | 4256.2 | 346.1 | 301.4 | 295.4 | 1156.5 | 1469.2 | 1755.7 | 3026.4 | 3654.5 | 3245.1 |
| Baah, 2011 | 4265.2 | 3156.4 | 4566.4 | 284.5 | 279.5 | 315.3 | 1254.4 | 1405.4 | 1821.6 | 2945.1 | 3824.4 | 2942.2 |

Table 7  Average number of statements examined to locate the first bug (worst case)

| Technique | Gzip | | | | Grep | | | | Sed | | | |
|---|---|---|---|---|---|---|---|---|---|---|---|---|
| | 3-bug | 4-bug | 5-bug | 6-bug | 3-bug | 4-bug | 5-bug | 6-bug | 3-bug | 4-bug | 5-bug | 6-bug |
| FPA-FL | **74.46** | **79.48** | **89.59** | **101.18** | 238.21 | **249.45** | 308.45 | **134.94** | **112.69** | 84.82 | 106.10 | 87.41 |
| D-Star | 120.49 | 98.25 | 182.5 | 188.45 | 255.67 | 281.32 | **290.76** | 150.33 | 144.62 | 111.62 | 84.60 | 85.16 |
| H3b | 109.26 | 120.88 | 135.51 | 115.96 | 371.26 | 378.38 | 391.62 | 142.88 | 288.21 | 428.12 | 79.30 | **81.52** |
| H3c | 129.26 | 141.85 | 158.1 | 119.42 | 399.19 | 391.36 | 380.56 | 146.12 | 224.51 | 361.17 | **75.26** | 87.18 |
| Ochiai | 165.88 | 185.25 | 215.48 | 202.46 | **226.78** | 341.26 | 378.96 | 176.61 | 351.65 | **79.45** | 376.96 | 525.6 |
| O | 1529.1 | 1668.4 | 1679.58 | 1511.6 | 421.65 | 391.85 | 394.62 | 295.71 | 2056.5 | 2128.8 | 2346.8 | 3190.4 |
| O<sup>P</sup> | 289.54 | 295.8 | 261.48 | 242.38 | 421.65 | 391.85 | 394.62 | 295.71 | 618.51 | 519.79 | 576.52 | 701.63 |
| Baah, 2011 | 144.62 | 208.30 | 194.72 | 216.48 | 234.92 | 312.66 | 354.35 | 148.36 | 246.18 | 135.6 | 212.35 | 154.86 |

Table 7  continued

| Technique | Make | | | | Space | | | | Bash | | | |
|---|---|---|---|---|---|---|---|---|---|---|---|---|
| | 3-bug | 4-bug | 5-bug | 6-bug | 3-bug | 4-bug | 5-bug | 6-bug | 3-bug | 4-bug | 5-bug | 6-bug |
| FPA-FL | **231.62** | 184.48 | 267.16 | 138.36 | **231.62** | 184.48 | 267.16 | 138.36 | **136.78** | **186.35** | **158.85** | **162.25** |
| D-Star | 271.50 | **179.36** | 297.98 | 160.33 | 271.50 | **179.36** | 297.98 | 160.33 | 169.65 | 279.15 | 235.25 | 196.44 |
| H3b | 461.8 | 490.50 | 478.45 | 198.12 | 461.8 | 490.50 | 478.45 | 198.12 | 325.75 | 366.75 | 421.63 | 256.26 |
| H3c | 660.12 | 621.56 | 669.6 | 265.11 | 660.12 | 621.56 | 669.6 | 265.11 | 397.39 | 385.33 | 452.72 | 295.15 |
| Ochiai | 291.28 | 309.53 | **252.89** | **132.21** | 291.28 | 309.53 | **252.89** | **132.21** | 246.16 | 305.42 | 272.55 | 205.75 |
| O | 5128.5 | 4911.6 | 4571.1 | 3651.3 | 5128.5 | 4911.6 | 4571.1 | 3651.3 | 288.82 | 295.89 | 311.40 | 186.39 |
| O<sup>P</sup> | 669.21 | 601.74 | 659.25 | 281.63 | 669.21 | 601.74 | 659.25 | 281.63 | 252.45 | 344.46 | 352.35 | 216.46 |
| Baah, 2011 | 348.54 | 296.22 | 318.46 | 186.36 | 348.54 | 296.22 | 318.46 | 186.36 | 212.62 | 226.80 | 238.26 | 185.58 |

Table 7  continued

| Technique | J-Chart | | | J-Time | | | C-Lang | | | C-Math | | |
|---|---|---|---|---|---|---|---|---|---|---|---|---|
| | 3-bug | 4-bug | 5-bug | 3-bug | 4-bug | 5-bug | 3-bug | 4-bug | 5-bug | 3-bug | 4-bug | 5-bug |
| FPA-FL | **4292.5** | **3366.7** | **4724.6** | **257.1** | **291.7** | **259.5** | **1288.4** | **1862.2** | **1969.8** | **3241.7** | **3624.6** | **3168.3** |
| D-Star | 4926.6 | 3859.7 | 5694.9 | 324.2 | 349.6 | 321.4 | 1868.7 | 2482.4 | 2526.9 | 3930.8 | 4231.4 | 3522.1 |
| H3b | 5152.8 | 4162.6 | 5324.6 | 395.8 | 334.4 | 345.2 | 2122.1 | 2652.5 | 3105.7 | 4181.7 | 4122.7 | 3937.4 |
| H3c | 4525.2 | 4344.4 | 5461.5 | 414.4 | 381.2 | 355.4 | 2552.6 | 3510.1 | 2644.3 | 4392.6 | 4661.3 | 3866.2 |
| Ochiai | 4652.9 | 3726.2 | 5521.5 | 331.6 | 355.8 | 305.3 | 1914.9 | 2345.8 | 2894.5 | 3944.3 | 5177.4 | 3441.7 |
| O | 5062.8 | 3566.7 | 5360.4 | 401.6 | 398.4 | 342.8 | 2458.4 | 2462.1 | 3250.6 | 4275.2 | 5235.8 | 3912.6 |
| O<sup>P</sup> | 5325.3 | 3752.6 | 5456.1 | 396.2 | 421.4 | 295.4 | 2656.3 | 2358.4 | 3574.2 | 3863.1 | 4381.2 | 4012.2 |
| Baah, 2011 | 4865.4 | 3856.4 | 5266.2 | 334.7 | 384.2 | 315.3 | 1854.2 | 2531.5 | 2592.7 | 3722.4 | 4610.1 | 3581.3 |

EXAM score-based comparison between FPA-FL and other peer techniques on Gzip, Grep and Defects4J programs are presented in Figure 6. From Part (a1), for Gzip, by examining 4 % of the code *FPA-FL* can locate the first bug of 85 (70.83%) of 120 multi-bug versions in the best cases and 77 (64.16 %) in the worst; H3b can locate 67 (55.83 %) for best cases and 52 (43.33 %) for the worst; Ochiai can get 55 (45.83 %) for best cases and 46 (38.33 %) for the worst; D-Star can locate 65 (54.16 %) for best cases and 50 (41.66 %) for the worst.

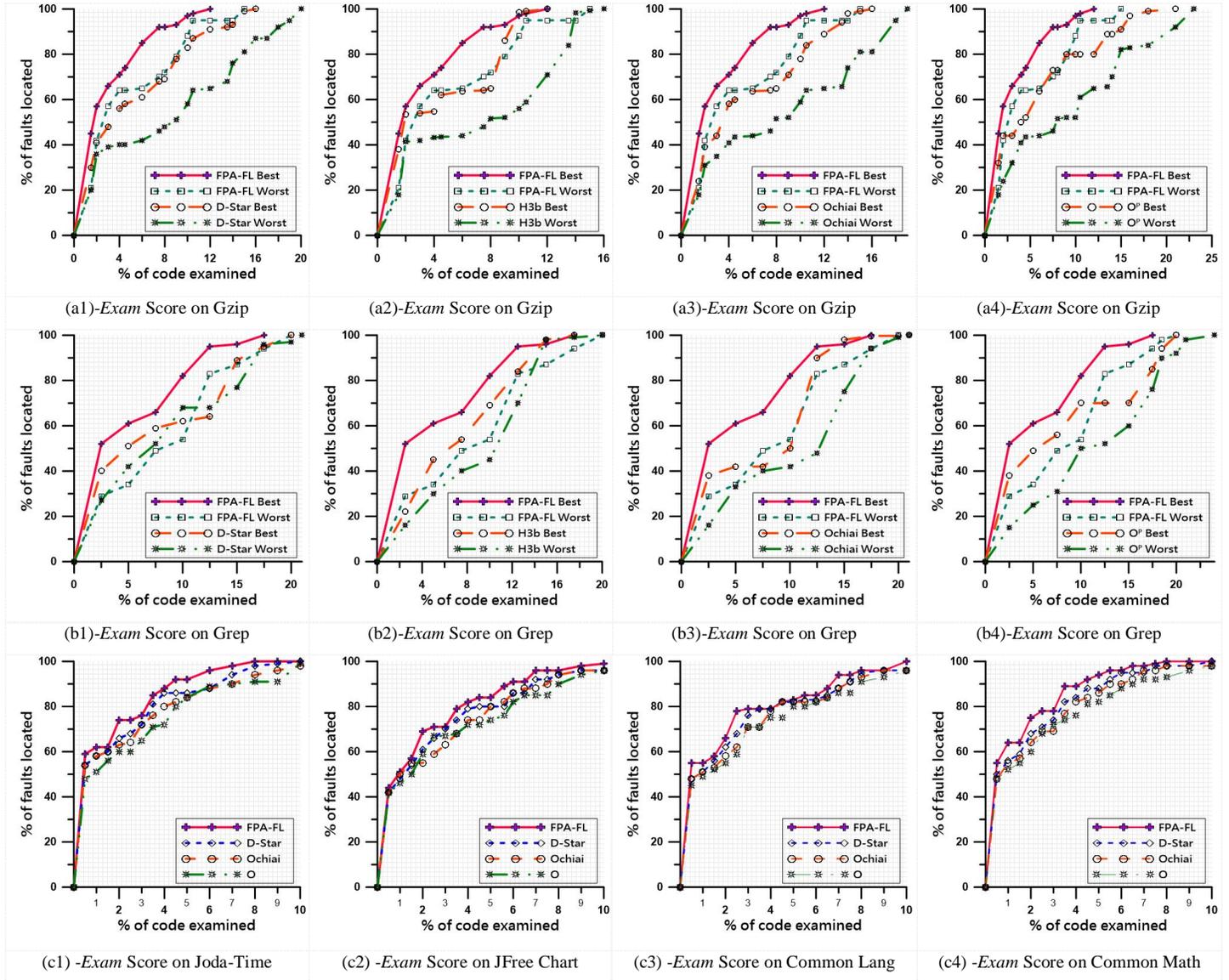

Figure 6. *EXAM* score-based comparison on Gzip, Grep and Defects4J programs (Multiple-bug case)

The grouping effect of the Elastic-Net method helps programmers detect multiple bugs in programs. FPA-FL puts highly correlated statements into a single group and attempts to construct a cause-effect chain for those statements. Therefore, the fault relevant statements corresponding to each individual bug are included into a group separated from groups including faulty statements of other existing bugs of a program.

## 3.5  Case study 3: Impact of static fault-proneness analysis and proposed slice coverage spectrum

In order to study the impact of considering the impact of static fault-proneness analysis in the Elastic-Net regression model, two cases are considered. First, the proposed model is constructed regardless of static fault-proneness analysis and with a uniform shrinkage parameter for all program statements. We named this model FPA-FL (uniform). Second, shrinkage parameter for each program statement is considered regarding its static FPL. In other words, in the second case, program statements are not treated equally, and a smaller shrinkage value is considered for fault-prone parts of programs. This model is called FPA-FL (Fault-prone). We then evaluate both models on Siemens, Gzip, Grep, Sed, Space and Make with five percent manual code inspection and for larger programs with one percent. Results are shown in Figure 7.

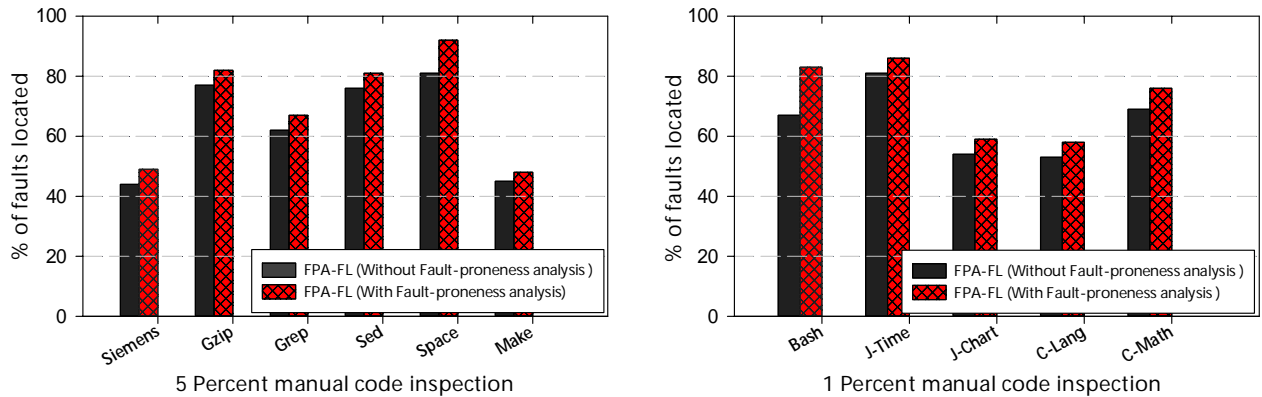

Figure 7. Performance of FPA-FL regarding to fault-proneness consideration

As we can see, FPA-FL (Fault-prone) performs better than FPA-FL (uniform) in localizing faults. As stated before, it is because FPA-FL (Fault-prone) puts more emphasis on fault-prone parts of programs. Put it differently, if a fault is located in complex parts of a program, FPA-FL (Fault-prone) will less likely omit the fault relevant statements from the regression model in comparison with FPA-FL (uniform).

We conducted another experiment to make an evaluation of the proposed slice coverage spectrum by comparing our approach with other fault localization methods. Tables 8 and 9 present the average number of statements that need to be examined by each fault localization technique across each of the subject programs, for both best and worst cases.

Table 8 Average number of statements examined with respect to each faulty version (Best case)

| Technique | Siemens | Gzip | Grep | Sed | Space | Bash | Make | J-Chart | J-Time | C-Lang | C-Math |
|---|---|---|---|---|---|---|---|---|---|---|---|
| FPA-FL | **9.21** | **51.2** | **102.94** | 66.32 | 35.90 | 76.33 | **115.84** | 4120.18 | 177.62 | **1238.48** | 2782.54 |
| D-Star | 12.35 | 54.25 | 125.64 | 70.45 | 37.15 | 87.63 | 248.76 | 4551.65 | 201.15 | 1336.56 | 2912.85 |
| H3c | 11.29 | 64.56 | 112.42 | 102.62 | 50.68 | 112.28 | 219.42 | 4580.26 | 192.69 | 1341.41 | 2840.62 |
| Ochiai | 13.77 | 61.72 | 141.22 | 69.45 | 37.45 | **71.24** | 246.63 | 4465.42 | 211.85 | 1356.25 | 2944.35 |
| O | 9.52 | 52.15 | 110.34 | **64.45** | **35.60** | 85.62 | 161.58 | 4792.14 | 285.64 | 1389.62 | 3612.24 |
| Baah, 2011 | 10.74 | 57.13 | 121.70 | 99.54 | 39.42 | 79.41 | 124.11 | 4392.68 | 199.44 | 1316.74 | 2826.71 |

Table 9 Average number of statements examined with respect to each faulty version (Worst case)

| Technique | Siemens | Gzip | Grep | Sed | Space | Bash | Make | J-Chart | J-Time | C-Lang | C-Math |
|---|---|---|---|---|---|---|---|---|---|---|---|
| FPA-FL | **18.85** | **119.52** | **191.45** | 176.26 | 63.41 | **106.75** | **248.36** | 4851.55 | **246.47** | **1424.28** | **3562.80** |
| D-Star | 19.21 | 121.52 | 212.35 | 174.81 | 75.25 | 118.14 | 416.5 | 5321.26 | 259.52 | 1572.92 | 3726.14 |
| H3c | 18.93 | 140.31 | 224.52 | 209.26 | 86.46 | 174.63 | 394.24 | 4996.52 | 268.48 | 1569.36 | 3618.80 |
| Ochiai | 20.85 | 128.46 | 236.36 | 187.45 | 70.33 | 124.25 | 411.52 | 5299.48 | 300.66 | 1580.41 | 3715.38 |
| O | 27.45 | 125.63 | 202.07 | **169.40** | **60.26** | 108.45 | 335.32 | 5416.62 | 346.89 | 1612.50 | 4025.92 |
| Baah, 2011 | 20.11 | 130.77 | 213.66 | 199.50 | 63.26 | 126.84 | 266.84 | 5027.37 | 288.47 | 1489.76 | 3602.45 |

For a detailed comparison, we adopt relative improvement (referred as Imp [56]). The Imp is to compare the total number of statements that need to be examined to find all faults using the slice coverage spectrum proposed in FPA-FL versus the number that needs to be examined by using the spectrum used in conventional SBFL. A lower value of Imp shows a better improvement over SBFL.

Table 10  Relative improvement of proposed slice coverage spectrum compared to the spectrum used in conventional SBFL (Best case)

| Technique | Siemens | Gzip | Grep | Sed | Space | Bash | Make | J-Chart | J-Time | C-Lang | C-Math |
|---|---|---|---|---|---|---|---|---|---|---|---|
| D-Star | 84 % | 92 % | 85 % | 88 % | 86 % | 94 % | 91 % | 94 % | 84 % | 94 % | 96 % |
| H3c | 86 % | 84 % | 78 % | 83 % | 86 % | 92 % | 89 % | 94 % | 83 % | 95 % | 97 % |
| Ochiai | 88 % | 90 % | 85 % | 85 % | 84 % | 95 % | 88 % | 93 % | 87 % | 93 % | 96 % |
| O | 93 % | 92 % | 88 % | 92 % | 95 % | 92 % | 85 % | 89 % | 83 % | 94 % | 87 % |
| Baah, 2011 | 84 % | 93 % | 89 % | 83 % | 87 % | 93 % | 88 % | 95 % | 88 % | 94 % | 96 % |

Table 11  Relative improvement of proposed slice coverage spectrum compared to the spectrum used in conventional SBFL (Worst case)

| Technique | Siemens | Gzip | Grep | Sed | Space | Bash | Make | J-Chart | J-Time | C-Lang | C-Math |
|---|---|---|---|---|---|---|---|---|---|---|---|
| D-Star | 84 % | 96 % | 94 % | 96 % | 91 % | 93 % | 92 % | 94 % | 88 % | 93 % | 94 % |
| H3c | 95 % | 95 % | 95 % | 91 % | 93 % | 89 % | 93 % | 97 % | 95 % | 93 % | 94 % |
| Ochiai | 88 % | 93 % | 92 % | 94 % | 94 % | 90 % | 90 % | 93 % | 85 % | 92 % | 92 % |
| O | 85 % | 95 % | 94 % | 97 % | 95 % | 93 % | 91 % | 92 % | 88 % | 94 % | 83 % |
| Baah, 2011 | 85 % | 93 % | 90 % | 94 % | 92 % | 90 % | 90 % | 93 % | 87 % | 94 % | 92 % |

As shown in Tables 10 and 11 the values of Imp on each program are less than 100%. This indicates that the proposed spectrum based on EBDS improves the effectiveness of statistical fault localization on all programs.

### 3.6  Complexity Analysis

Finally, we present the time overhead of FPA-FL comparing with Ochiai, as a representative of SBFL approaches in Table 12. Column *preprocessing* illustrates the time overhead spent for all test cases by fault localization approaches, including the time required for computing slices for *FPA-FL*. Column *Static FPL Estimation & Clustering* presents the computational time of estimating the static fault-proneness likelihood of program statements and clustering execution vectors for identifying likely coincidentally correct tests. Column *suspiciousness computing* presents the computational time of suspiciousness, and the last column presents the division of the total time required by *FPA-FL* and Ochiai to illustrate the high time cost of our approach regarding to Ochiai. Column $\#FPA-FL/\#Ochiai$ in Table 12 shows that the time overhead of *FPA-FL* for all subjects are less than 1.94 times of Ochiai in our experiment.

Table 12  Time costs of SBFL and FPA-FL spent on all subjects

| Subject | Approach | Preprocessing | Static FPL Estimation & Clustering | Suspiciousness Computing | Total | $\frac{\#FPA-FL}{\#Ochiai}$ |
|---|---|---|---|---|---|---|
| Siemens | FPA-FL | 49,625s | 11845s | 4,936s | 66,406s | 1.67 |
| | Ochiai | 36,826s | 0 | 2,724s | 39,550s | |
| Gzip | FPA-FL | 5,093s | 2121s | 821s | 8,035s | 1.94 |
| | Ochiai | 3,926s | 0 | 211s | 4,137s | |
| Grep | FPA-FL | 6,236s | 1883s | 1,066s | 9,185s | 1.88 |
| | Ochiai | 4,394s | 0 | 476s | 4,870s | |
| Sed | FPA-FL | 5,614s | 1565s | 962s | 8,141s | 1.83 |
| | Ochiai | 4,027s | 0 | 421s | 4,448s | |
| Space | FPA-FL | 28,531s | 4241s | 1263s | 34,035s | 1.40 |
| | Ochiai | 23,852s | 0 | 326s | 24,178s | |
| Make | FPA-FL | 11,246s | 2766s | 941s | 14,953s | 1.72 |
| | Ochiai | 8,268s | 0 | 389s | 8,657s | |
| Bash | FPA-FL | 5,849s | 2194s | 630s | 8,673s | 1.79 |
| | Ochiai | 4,681s | 0 | 156s | 4,837s | |
| J-Time | FPA-FL | 37,531s | 7665s | 3,962s | 49,158s | 1.65 |
| | Ochiai | 28,852s | 0 | 850s | 29,702s | |
| J-Chart | FPA-FL | 32,531s | 4741s | 4,322s | 41,594s | 1.60 |
| | Ochiai | 24,852s | 0 | 1026s | 25,878s | |
| C-Lang | FPA-FL | 46,531s | 11762s | 3,156s | 61,449s | 1.54 |
| | Ochiai | 38,852s | 0 | 889s | 39,739s | |
| C-Math | FPA-FL | 53,531s | 14494s | 3,520s | 71,545s | 1.58 |
| | Ochiai | 43,852s | 0 | 1231s | 45,083s | |

### 4.  Discussion and Related Work

In this section, we briefly review the previous works related to fault localization in general. The idea of using a regression method for fault localization was first introduced by Dickinson et al. in [57] and developed by Podgurski et al. in [58]. The method in [57] uses a naïve logistic regression to select suspicious features from program runtime profiles and clusters the selected features to identify similar features. Liblit et al. [59] use a regularized logistic regression to select suspicious predicates which are correlated with software crash. Liblit's method contains some deficiencies. Since in large programs with a huge number of predicates, the majority of predicates are redundant and irrelevant to the failure, the regression model in [59] may retain some irrelevant predicates in the model. The main problem with the logistic regression in [59] is that it is not appropriate for programs with a high amount of correlation among predicates. Furthermore, it is incapable to find multiple bugs. Since the traditional regression method is inadequate for applications with highly correlated variables, in Parsa et al. [60], a ridge regression model has been applied. They also introduce a combination of ridge and lasso regression methods in [58]. The ridge regression is helpful when there is high correlation among predicates. However, for large programs with huge numbers of predicates, it could not provide interpretable models. The lasso method has the feature selection capability that removes irrelevant features and preserves the significant ones. However, when there is high correlation among predicates, it assigns relatively high coefficient to a single predicate and very small coefficient to the other correlated ones. Furthermore, it does not have grouping effect. Both ridge and lasso methods fail to work for $P \gg N$ problems where the number of predicates is much more than the number of executions [25]. This may limit their scalability for large programs and small input data (i.e., test cases). In [25] the Elastic-Net regression is applied on program predicates by Parsa et al. Elastic-Net penalty parameter is a combination of both ridge regression and lasso penalties, and therefore, it has the strong points of both techniques. It contains the grouping effect and is capable of dealing with the $P \gg N$ problem to some extent. FPA-FL technique resolves the mentioned difficulties and provides precise interpretable models even from very large programs including highly correlated predicates. It also works accurately when the number of test cases is reasonably small. This paper extends our previous works presented in [3][25-26] in some important ways. FPA-FL takes into consideration the fault-proneness of program statements which calculated based on complexity code metrics. The previous works did not address the multiple-bug problem which is widely studied in this paper by evaluating FPA-FL on multiple-bug versions. Finding the cause-effect chains as the failure context is another advantage of FPA-FL which was not considered in our previous works.

Recently Baah et al. used causal inference and regression modeling in order to take into consideration multiple statements in their fault localization analysis [52-53]. Given a statement *s* in program *P*, they obtained a causal-effect estimate of *s* on the outcome of *P* that is not subject to severe confounding bias, i.e., a causation-based suspiciousness score of s that takes into consideration other statements that relate to *s* via control/data dependence. They estimated the suspiciousness score of *s* based on a linear regression model. Since Baah et al.'s techniques [52-53] involve fitting a regression model or computing a matching for each statement in a program, their computational and profile storage overheads may be considerable for large programs and large test suites. Therefore, they suffer from scalability issues. On the contrary, scalability is one of the strengths of our method. Unlike Baah et al.'s techniques, FPA-FL, builds only a single regularized logistic regression model based on fault candidate program statements. Another advantage of FPA-FL in comparison with Baah et al. approaches is the grouping effect of Elastic-Net regression. Using Elastic-Net regression, program statements are grouped according to their impact on program termination state and the groups are distinct and recognizable from each other.

Context aware technique [62] is another related work that proposed by Jiang et al. It combines the methods of feature selection, clustering, and static control flow graph analysis to identify the failure context. It provides two types of information: (1) Identifying fault suspicious predicates and (2) recognizing correlated predicates. To identify the predicates which might be relevant to the program failure, it applies support vector machine and random forest techniques to perform feature selection. To find the correlated predicates, it applies the k-means clustering method. These two types of information are used to predict the direction of conditional branch statements which lead a program to failure. To study the combinatorial effect of predicates on program failure, Arumuga et al. in [63] combined the simple boolean expressions into compound predicates The combination is done according to the data/control dependence among predicates in a program dependence graph. They applied logic operators to make complex predicates and estimated their impact on program failure using some conditional probabilities. Since the predicates are combined artificially, some of these complex predicates may not actually exist in any execution. Furthermore, for large programs with many predicates, the huge number of complex predicates can cause scalability issues.

A program spectrum details the execution information of a program from certain perspectives and can be used to track program behavior. SBFL techniques use program spectrum to indicate entities more likely to be faulty. The Ochiai similarity coefficient [4] has also been used in the context of software fault localization. It, based on the same heuristic as Tarantula [46]. According to the data reported in [4], Ochiai is more effective in fault localization than Tarantula. Wong et al., [5] propose a method called DStar that modifies a similarity-based coefficient (i.e., Kulczynski coefficient) to better localize the bugs as compared to Ochiai, and 12 similarity-based coefficients. Since we cannot theoretically prove that other values of its exponent shall always be more efficient in fault localization, without loss of generality, the exponent of D-Star is set to 2 for simplified calculation in our empirical study. In [6], Naish et al. propose two techniques, $O$ and $O^P$. The technique $O$ is designed for programs with a single bug, while $O^P$ is better applied to programs with multiple bugs. Data from their experiments suggest that $O$ and $O^P$ are more effective than Tarantula and Ochiai for single-bug programs. A fault localization technique based on crosstab (cross-tabulation) which has been used to study the relationship between two or more categorical variables is proposed in [8]. Crosstabs constructed by two column-wise categorical variables (covered and not covered) and two row-wise categorical variables (passing and failing) are used to aid in analyzing the association degree between the failing (or passing) execution result and the coverage of each statement with null hypothesis that they are independent given beforehand. In addition, the degree of association will be used to determine the suspiciousness value for each statement. A unique aspect of this technique is that a well-defined statistical analysis is applied to locating faults in software. The impact of how each additional failing (or passing) test case can help locate program faults is investigated in [7]. The conclusion is that the contribution of the identified failing test cases are stepwise diminishing such as the first failing test contributes larger than or equal to that of the second failing test, which is larger than or equal to that of the third failing test. This conclusion also holds for passing tests. The major difference between H3b and H3c and others is that they emphasize that the contribution of each additional failing and passing test in computing the suspiciousness of each statement should be different, whereas others assume an equal contribution for every failing and passing test. Comparisons among different SBFL techniques are frequently discussed in recent studies [5-6]. However, there is no technique claiming that it can outperform all others under every scenario.

Most of the earlier debugging effort relied on slicing-based techniques to help programmers reduce the search domain to quickly locate bugs [64-66]. In [64], Xiaolin et al. proposed a method using a hybrid spectrum of full slices and execution slices to improve the effectiveness of fault localization. Their approach firstly computes full slices of failed test cases and execution slices of passed test cases respectively. Secondly it constructs the hybrid spectrum by intersecting full slices and execution slices. Finally, it computes the suspiciousness of each statement in the hybrid slice spectrum and generates a fault location report with descending suspiciousness of each statement. Agrawal et al. [67] apply the execution slice to fault localization by examining the execution dice of one failed and one successful test to locate program bugs. A program dice is defined as the set of statements that affect the computation of incorrect variable while do not affect the computation of the correct one.

Dynamic slicing can be performed in three different ways: Data slicing, full slicing and relevant slicing [28]. Data slicing computes slices by traversing a dynamic data dependence graph. Full slicing considers both data and control dependence in computing dynamic slices. Therefore, all statements which have direct or indirect impact on faulty output value through dynamic data or/and control dependences are included in the slice. A relevant slice contains both statements in a full slice and relevant statements. Relevant statements are those statements which do not have actual influence on output value but they could have a potential to affect the output value if they have been evaluated in a different way [28]. Conducted researches show that the relevant slicing could capture more faults compared with full slicing which in turn could be more powerful than data slicing. However, it may contain irrelevant static data and thereby we proffered to compute full slicing in our proposed approach and expand the computed slice with statements that could have a potential to affect the output value. Although it has been shown that dynamic slicing substantially reduces the number of program statements to be examined, the absolute number of statements might still be large and many of the statements in the slice are unlikely to be fault relevant [27][68].To reduce the size of dynamic slices, Zhang et al. in [69] integrate dynamic backward slicing with the idea of delta debugging [70]. In this regard, they first identify a minimal failure-inducing input and then compute the forward slice starting from the failure-inducing input. The computed forward slice is finally intersected with the backward dynamic slice of an incorrect output. In a later work [71], they compute a bidirectional dynamic slice for critical predicates. The critical predicate is a conditional branch which is likely to be responsible for the failure execution and if an execution instance of that predicate is switched from one outcome to another, the program produces a desirable output. The bidirectional dynamic slice contains statements in both forward and backward slices of a critical predicate. In a similar work by Jeffrey et al. [68], a set of values which have been used in an execution instance of a statement in a failing execution is replaced with some other values to analyze whether the program result changes from incorrect to correct output. If that happens, the statement is marked as a faulty statement for a single bug programs.

Aforementioned slice-based methods have some limitations. None of them are able to rank statements according to their fault relevance and thereby human debugger should examine all the reported statements with the same chance to find failure origin. Moreover, a minimal failure-inducing input cannot be computed for all type of programs. In many situations, all input parameters may have impact on program incorrect result and cannot be simplified and isolated any further. On the other hand, techniques that rely on critical predicate switching [71] or value replacement [68] may have scalability problems for large programs with huge amount of data values. In predicate switching case, there might be many predicates in a program and a failure output may be the cause of more than a single critical predicate [63]. Thereby, identifying critical predicates among huge number of predicates does not seem to be trivial and straightforward. Another limitation of these techniques is their dependence on a single failing execution. Hence, they only identify a fault that is related to that failure and cannot detect other unknown faults of a program. They are also incapable of finding multiple bugs.

## 5. Threats to Validity

Similar to any empirical study, there are some threats to the validity of our experiments. Threats to external validity arise when the results of the experiment are unable to be generalized to other situations. Siemens suite contains small-sized programs which by their own cannot justify our claim about the performance of FPA-FL. To emphasize the effectiveness of FPA-FL, we have also applied our technique on Gzip, Grep, Sed, Make, Space and Bash programs which are larger programs with real-life and more complex bugs. Each of these programs varies greatly from the other with respect to size, function, number of faulty versions studied, etc. Recent studies have shown that the fault localization results using real faults are different from those using seeded faults. They found that artificial faults are not useful for predicting which fault localization techniques perform best on real faults. Thus, we also evaluated FPA-FL on Defects4J suite which consists of 224 real faults from 4 open source projects. This allows us to have greater confidence in the applicability of FPA-FL method to different programs and the superiority of the results. However, the nature of faults in code, in general, is highly diverse and complex, and therefore our results may not be representative for all possible programs. A potential threat to the construct validity of debugging methods is the adequacy of quality metric that is chosen to measure their reported results. To mitigate this threat, we use evaluation metrics which are widely used in the literature of software fault localization. However, these metrics may not reflect the real-world cost of manually localizing the faults since, for example, for the results presented in this paper, we assume that if a programmer examines a faulty statement, he/she will identify the corresponding fault. At the same time, a programmer will not identify a non-faulty statement as faulty. If such perfect bug detection does not hold, then the number of statements that need to be examined to identify a bug may increase. However, such a concern also applies to other fault-localization techniques, and therefore, this is a common limitation. Furthermore, some categories of faults are easier to be located than the others. For instance, a missing code fault is relatively difficult to be identified in comparison with a fault which is located in an assignment statement. The use of a test suite with a different size may give a different result [72]. Evaluations on the impact of different test suite sizes on our technique are performed on Bash program which is a considerably large program. However, more evaluations would be welcome.

## 6. Conclusion

In this article a new combinatorial approach to fault localization, so-called FPA-FL is proposed. It considers both the dynamic and static attributes of programs to rank the statements according to their effects on the program termination state. We attempt to alleviate the bias introduced by the test data through taking into account the static structure of a program while modeling its dynamic behavior. To minimize the search space, the statements appearing in expanded dynamic slices are initially selected for the ranking. Using a combination of the statistical approach to fault localization and backward slicing, we take advantage of the capability of statistical approaches to rank statements and the strength of slicing in restricting the statements to failure relevant ones. We claim that the most program failures are only revealed when a specific combination of correlated statements are executed. Hence, studying the combinatorial effect of correlated program statements on the program termination state is of great importance. We depict that the Elastic-Net regression method can be best applied to address this problem. Fault localization algorithms are mainly evaluated by measuring the amount of code to be manually examined around their reported fault suspicious statements before the actual failure origin is located. The grouping effect of the fault suspicious statements could reveal multiple bugs in a program. The claim is supported by evaluating FPA-FL on multiple-bug versions of Siemens, Defects4J, Gzip, Grep, Sed, Space, Make and Bash programs. FPA-FL also finds the failure context by identifying cause-effect chains. The grouping effect of the Elastic-Net technique makes FPA-FL highly scalable as can be seen by the promising results on the subject programs.